\documentclass
[12pt]{article}
\usepackage[dvips]{color}
\usepackage{epsfig}
\usepackage{amssymb,eucal}
\usepackage{amsmath}
\usepackage{graphicx}
\newcommand{\Mkk}{M_{\rm KK}}
\newcommand{\Nc}{N_{c}}
\newcommand{\Nf}{N_{f}}

\begin{document}
\title {Application of AdS/CFT in Nuclear Physics}
\author{ M. R. Pahlavani\thanks{Email: m.pahlavani@umz.ac.ir}\hspace{1mm}\\
{\small {\em  Department of Physics, Faculty of Sciences, University of Mazandaran, }}\\
{\small {\em P. O. Box 47415-416, Babolsar, Iran}}\\
 R. Morad\thanks{Email: r.morad@umz.ac.ir}\hspace{1mm}\\
{\small {\em Department of Physics, University of Cape Town, Private Bag X3, Rondebosch 7701, South Africa}}}
 \maketitle
\begin{abstract}

We review some recent progress in studying the nuclear physics especially
nucleon-nucleon (NN) force within the gauge-gravity duality, in
context of noncritical string theory. Our main focus is on the
holographic QCD model based on the $AdS_6$ background. We explain the
noncritical holography model and obtain the vector-meson spectrum
and pion decay constant. Also, we study the NN interaction in this
frame and calculate the nucleon-meson coupling constants. A further
topic covered is a toy model for calculating the light nuclei
potential. In particular, we calculate the light nuclei binding
energies and also excited energies of some available excited states.
We compare our results with the results of other nuclear models and
also with the experimental data. Moreover, we describe some other
issues which are studied using the gauge-gravity duality.


\end{abstract}

\tableofcontents

\maketitle \nonstopmode

\section{Introduction}

One of the fundamental ingredients of nuclear physics is the nuclear
force with which point-like nucleons interact with each other. Since
Yukawa, many potential models have been constructed which have been
composed to fit the available NN scattering data. The newer
potentials have only slightly improved with respect to the previous
ones in describing the recent much more accurate data. As it is
shown in Ref[1], all of these potential models do not have good
quality with respect to the pp scattering data below 350 MeV and
just a few of them are of satisfactory quality. These models are the
Reid soft-core potential Reid68 [2], the Nijmegen soft-core
potential Nijm78 [3], the new Bonn pp potential Bonn89 [4] and also
the parameterized Paris potential Paris80 [5]. These familiar
one-boson-exchange potentials (OBEP) contain a relatively small
number of free parameters (about 10 to 15 parameters), but do not
have a reasonable description of the empirical scattering data.
Also, most of these potentials which have been fitted to the np
scattering data, unfortunately do not automatically fit to the pp
scattering data even by considering the correction term for the
Coulomb interaction [1]. Of course, new versions of these potentials
have been constructed such as Nijm I, Nijm II, Reid93 [6], CD-Bonn
[7], and AV18 [8] which explain the empirical scattering data
successfully. But they contain a large number of purely
phenomenological parameters. For example, an updated (Nijm92pp [9])
version of the Nijm78 potential contains 39 free parameters.

On the other hand, there are many attempts to impose the symmetries
of QCD using an effective Lagrangian of pions and nucleons [10,11].
These models only capture the qualitative features of the nuclear
interactions and could not compete with the much more successful
potential models mentioned above.

Despite many efforts, no potential model has yet been constructed which gives a high-quality description of the empirical data, obeys the symmetries of QCD, and contains only a few number of free phenomenological parameters.

In recent years, holography or gauge-gravity duality gave us a new approach to hadronic physics [12] and make new progress in understanding the nuclear force.

Nuclear force, the force between nucleons, exhibits a repulsive core of nucleons at short distances. This repulsive core is quite important for large varieties of physics of nuclei and nuclear matter. For example, the well-known presence of nuclear saturation density is essentially due to this repulsive core. However, from the viewpoint of strongly coupled QCD, the physical origin of this repulsive core has not been well understood. Nuclear force especially the repulsive core has been studied using the AdS/CFT correspondence [13-16] and an explicit expression has been obtained for the repulsive core.

Also, there are many attempts to find a geometry dual to nucleus. Since nucleons are described by D-branes wrapping a sphere in curved geometry of holographic QCD, on a nucleus with mass number A there appears a U(A) gauge theory. One can find the dual gravity By taking the large mass number limit $A\rightarrow\infty$ and obtained a near horizon geometry corresponding to the heavy nucleus. The corresponding supergravity solution has discrete fluctuation spectra  comparable with nuclear experimental data [17]. As we know from the nuclear experiments, the nucleons of a heavy nuclei have coherent excitations which are called Giant resonances. These resonances exhibit harmonic behavior $E_n = n w(A)$ which is explained with phenomenological models such as the liquid drop model. The gauge-gravity duality can reproduce this behavior. Moreover, dependence to the mass number A is obtained by using the duality [17].

Among the holographic QCD models, the Sakai-Sugimoto (SS) [18-19] and Klebanov-Strassler (KS) models [20] are the most interesting holographic models to study strong coupling regime of QCD. The SS model is based on ten-dimensional type-IIA string theory, with a background geometry given by $N_c$ D4-branes. They fill four-dimensional Minkowski space-time and extend along a fifth extra dimension $x_4$ compactified on a circle whose circumference is parametrized by the Kaluza-Klein mass. Through this compactified dimension and antisymmetric boundary conditions for fermions supersymmetry is completely broken. Left- and right-handed chiral fermions are introduced by adding $N_f$ D8- and $N_f$ D8-branes which extend in all dimensions except $x_4$. In this compact direction, they are separated by a distance $L\in[0, \pi / M_{KK} ]$. There are two possible background geometries called confined and deconfined phase. For more details about the setup of the model see the original papers by Sakai and Sugimoto, refs. [18-19]. In this model, there is a nice topological interpretation of chiral symmetry breaking.

Chiral symmetry breaking is realized in the model as follows. A U($N_f$ ) gauge symmetry
on the flavor branes corresponds to a global U($N_f$ ) at the boundary. Therefore, the bulk
gauge symmetries on the D8- and D8-branes can be interpreted as left- and right-handed
flavor symmetry groups in the dual field theory. The Chern-Simons term accounts for the
axial anomaly of QCD, such that one is left with the chiral group $SU (N_f )_L \times SU (N_f )_R$
and the vector part $U(1)_V$ . There is no explicit breaking of this group since the model
only contains massless quarks. Spontaneous chiral symmetry breaking is realized when
the D8- and D8-branes connect in the bulk. They always connect in the confined phase whether they connect in the deconfined
phase depends on the separation L of the D8- and D8-branes in the extra dimension $x_4$.

The Sakai-Sugimoto model is particularly suited for phenomenon related to chirality as chiral magnetic effect (CME) [21-25] since it has a well defined
concept for chirality and the chiral phase transition. It is straightforward to introduce right- and left-handed chemical potentials independently.

The chiral magnetic effect is a hypothetical phenomenon which states that, in the presence of a magnetic field B, a nonzero axial charge density will lead to an electric current along the direction of the B field [26-28]. Analysis of RHIC data appears to favor the presence of a CME in the quark-gluon plasma, although a better understanding of systematic errors and backgrounds is still needed. CME is studied in many holographic systems, following refs. [29-34], including systems
without confinement or chiral symmetry breaking in vacuum.

Also, predictions of the SS model are in good agreement with the lattice
simulations such as the glueball spectrum of pure QCD [35-36]. This model describes baryons and their interactions with mesons well [18-19,37-39]. It is shown that the baryons can be taken as point-like objects at distances larger than their sizes, so their
interactions can be described by the exchange of light particles
such as mesons. Therefore, one can find the baryon-baryon potential
from the Feynman diagrams using the interaction vertices including
baryon currents and light mesons [38]. But there are some
inconsistencies. For example, the size of the baryon is proportional
to $\lambda^{-1/2}$. Consequently in the large 't Hooft coupling
(large $\lambda$), the size of the baryon becomes zero and the
stringy corrections have to be taken into account. Another problem
is that the scale of the system associated with the baryonic
structure is roughly half the one needed to fit to the mesonic data
[40]. Also, the holographic models arising from the critical string
theory encounter with the some Kaluza-Klein (KK) modes, with the
mass scale of the same order as the masses of the hadronic modes.
These unwanted modes are coupled to the hadronic modes and there is
no mechanism to disentangle them from the hadronic modes yet. In
order to overcome this problem, it is possible to consider the color
brane configuration in non-critical string theory [41-44].

The non-critical string is not formulated with the critical dimension, but nonetheless has vanishing Weyl anomaly.
A worldsheet theory with the correct central charge can be constructed by introducing a non-trivial target space, commonly by giving an expectation value to the dilaton which varies linearly along some spacetime direction. For this reason non-critical string theory is sometimes called the linear dilaton theory. Since the dilaton is related to the string coupling constant, this theory contains a region where the coupling is weak (and so perturbation theory is valid) and another region where the theory is strongly coupled [46-47].

In such backgrounds the string coupling constant is proportional to
$\frac{1}{N_c}$, so the large $N_c$ limit corresponds to the small
string coupling constant. However, contrary to the critical
holographic models, in the large $N_c$ limit, the 't Hooft coupling
is of order one instead of infinity and the scalar curvature of the
gravitational background is also of order one. So, it seems the
non-critical gauge-gravity correspondence is not very reliable. But
studies show that the results of these models for some low energy
QCD properties such as the meson mass spectrum, Wilson loop, and the
mass spectrum of glueballs [45-47] are comparable with lattice
computations. Therefore non-critical holographic models still seem
useful to study QCD.

One of the non-critical holographic models is composed of a $D4$ and
anti $D4$ brane in six-dimensional non-critical string theory
[43,45]. The low energy effective theory on the intersecting brane
configuration is a four-dimensional QCD-like effective theory with
the global chiral symmetry $U(N_f)_L \times U(N_f)_R$. In this brane
configuration, the six-dimensional gravity background is the near
horizon geometry of the color $D4$ branes. This model is based on
the compactified $AdS_6$ space-time with constant dilaton. So the
model does not suffer from large string coupling as the SS
model. The meson spectrum [45] and the structure of thermal phase
[48] are studied in this model. Some properties, like the dependence
of the meson masses on the stringy mass of the quarks and the
excitation number are different from the critical holographic models
such as the SS model.

We study the gauge field and its mode expansion in this non-critical
holography model and obtain the effective pion action [49]. The model
has a mass scale $M_{KK}$ like the SS model in which we set its
value by computing the pion decay constant. Then, we study the
baryon [50] and obtain its size. We show that the size of the baryon
is of order one with respect to the 't Hooft coupling, so the
problem of the zero size of the baryon in the critical holography
model is solved. But the size of the baryon is still smaller than
the mass scale of holographic QCD, so we treat it as a point-like
object and introduce an isospin $1/2$ Dirac field for the baryon
[49]. We write a 5D effective action for the baryon field and reduce
it to 4D using the mode expansion of gauge field and baryon
field and obtain the NN potential in terms of the meson exchange
interactions. We calculate the meson-nucleon couplings using the
suitable overlapping wave function integrals and compare them with
the results of SS model. Also, we compare the nucleon-meson couplings obtained from noncritical holographic model with the results of SS model and predictions of some phenomenological models. Our study shows that the noncritical results are in good agreement with the other available models.

On the other hand, one of the oldest problems of nuclear physics is
the nuclear binding energies: The interactions between nucleons are
very strong, while the nuclear matter is not relativistic. Nuclear
binding energies are experimentally known with high accuracy while
they are not predicted with sufficient accuracy using different
theoretical models. Since, prediction of nuclear binding energy is a
useful tool to test the goodness of a theoretical nucleon-nucleon
(NN) interaction model, we use our NN holography potential to obtain
the light nuclei binding energies. We construct a nuclear
holographic model [50-53] in the noncritical base and calculate the
nuclei potentials as the sum of their NN interactions. The minimum
of the ground state potential is considered as the binding energy.
Also, difference between this energy and the minimum of the excited
state potential presents the excited energy for each state. In order
to compute the potentials, we use the values of nucleon-meson
coupling constants obtained from both the critical and noncritical
holography models.

This paper is organized as follows: In Sec. 2 we briefly review the
AdS/CFT correspondence. The noncritical holographic model is
introduced in Sec. 3 and NN potential is constructed in this
section. In Sec. 4 we construct a simple model to study light nuclei
such as $^{2}D$, $^{3}T$, $^{3}He$ and $^{4}He$ and obtain their
potential of ground and excited states and respective binding
energies. Section 6 is devoted to a brief summary and conclusions.
Also, some other topics which are studied using the duality, are
introduced in this section.


\section{Review of AdS/CFT Correspondence}
\subsection{Historical Notes}

Quantum Chromodynamics (QCD) is the quantum field theory of the
strong interactions which has two important properties, asymptotic
freedom and confinement. Various analytical and numerical methods
have been developed to study QCD. One example is perturbative QCD
which works at small distances where the coupling is weak, but fails
to work at larger distances where the coupling becomes relatively
strong in which case the problem is said to become non-perturbative.
Examples of methods that study non-perturbative problems are
effective field theories such as chiral perturbation theory, lattice
QCD [54], Dyson-Schwinger equations (DSE) formalism [55] and
gauge/gravity duality [12,56-57].

Before QCD, in the 1960's string theory was introduced as a model to
describe the strong interactions [58]. It was able to explain the
organization of hadrons in Regge trajectories, describing them as
rotating strings. After the formulation of QCD, string theory took a
different direction, becoming a possible candidate for a unified
theory of all the forces. Nevertheless, some string interpretation
of hadron spectra was not abandoned; for example, a meson is
sometimes described as a quark and an anti-quark connected by a tube
of strong interaction flux [59-60]. This picture establishes a link
between QCD and string theory, which becomes even more evident in
the limit of large number of colors N [61]. 't Hooft proposed that
in this limit the gauge theory may have a description in terms of a
tree level string theory; in particular, the leading Feynman
diagrams in the 1/N expansion are planar and look like the
worldsheet of a string theory. For example, a meson can be
represented by two quark lines propagating in time connected by a
dense "sheet" of gluons, reminding the worldsheet swept out by a
string through time. In 1997 these studies found a possible new
framework in the so-called AdS/CFT correspondence [12], a conjecture
introduced by Maldacena relating a supergravity theory in ten
dimensions to a supersymmetric gauge theory in four dimensions. This
correspondence has been extended to a gauge theory as $SU(3)_c$
,thus proving some link between QCD and a higher dimensional theory
in a curved space-time.

\subsection{D-branes and $AdS$ Space}

The most important property of D-branes is that they contain gauge
theories on their world volume. In particular, the massless spectrum
of open strings living on a Dp-brane contains a (maximally
supersymmetric) U(1) gauge theory in p + 1 dimensions. Moreover, it
appears that if we consider the stack of N coincident D-branes, then
there are $N^2$ different species of open strings which can begin
and end on any of the D-branes, allowing us to have (maximally
supersymmetric) U(N) gauge theory on the world-volume of these
D-branes. Now, if N is sufficiently large, then this stack of
D-branes is a heavy object embedded into a theory of closed strings
that contains gravity. This heavy object curves the space which can
then be described by some classical metric and other background
fields.

Thus, we have two absolutely different descriptions of the stack of
coincident Dp-branes. One description is in terms of the U(N)
supersymmetric gauge theory on the world volume of the Dp-branes,
and the other is in terms of the classical theory in some
gravitational background. It is this idea that lies at the basis of
gauge/gravity duality.

One important example is D3-branes which can also be seen as
solutions of ten dimensional type IIB supergravity at low energies,
with metric of the form [62],

\begin{equation}
ds^2 = \left( 1 + \frac{L^4}{r^4} \right)^{-1/2} \left[ -dt^2 +
d\vec x^2 \right] + \left( 1 + \frac{L^4}{r^4} \right)^{1/2} \left[
dr^2 + r^2\, d\Omega_5^2 \right]\, ,
\end{equation}
where
\begin{equation}
L^4 = 4 \pi g_s N^2 \alpha' ,
\end{equation}
here, $g_s$ is the string coupling constant which is related to the
constant dilaton as ($g_s = e^\Phi$). Also, there is $N_c$ units of
$F_{[5]}$ flux. $L$ is the only length scale in the solution. This
metric interpolates between a throat geometry and a ten dimensional
Minkowski region.

If we take the {near horizon limit} of the solution given in eq.
(1), $r \ll L$, and redefine $z = L^2/r$, we can completely decouple
the Minkowski region and are left with a throat geometry which is
given by
\begin{equation}
ds^2 = \frac{L^2}{z^2} \left[ -dt^2 + d\vec x^2 + dz^2 \right] +
L^2\, d\Omega_5^2 \, ,
\end{equation}
which is the Poincar\'e wedge of the direct product of five
dimensional anti-de-Sitter space and a five sphere (AdS$_5 \times$
S$^5$). The isometry group of this space is given by $SO(4,2) \times
SO(6)$, though if we include fermions, the full supersymmetric
isometry group is {$PSU(2,2|4)$}. Note that this is exactly the same
as the full global symmetry group of the low energy limit of the
open string sector ({\it i.e.} SYM theory).

We see that the radius $L$, of both the AdS throat and the $S^5$, in
string units is given in terms of the gauge theory parameters as

\begin{equation}
L^4= g^2_{YM}\, N_c\, {\alpha^\prime}^{2}=\lambda{\alpha'}^2 \,.
\end{equation}

Therefore, in order that the stringy modes be unimportant,
$L\gg\sqrt{\alpha'}$, which translates into gauge theory language as
$\lambda = g^2_{YM}\, N_c \gg 1$.

\subsection{$\cal N=$ 4 Super-Yang-Mills theory}

$\cal N =$ 4 $SU(N)$ supersymmetric Yang-Mills theory (SYM) in four
dimensions (the dimensionality of the world-volume of the D3-branes)
has one vector field, $A_\mu$, six scalars fields $\phi^I$
($I=1...6$), and four fermions $\chi_{\alpha}^i$,
$\chi_{\dot\alpha}^{\bar i}$ ( $i,\bar i=1,2,3,4$) which are in the
${\bf 4}$ and $\bar {\bf 4}$ representations of the  $SU(4) = SO(6)$
$R$-symmetry group.

This theory naturally arises on the surface of a D3 brane in type
IIB superstring theory. Open strings generate a massless gauge field
in ten dimensions. When the open string ends are restricted to a 3+1
dimensional subspace the ten components of the gauge field naturally
break into a 3+1 dimensional gauge field and 6 scalar fields. The
fermionic super-partners naturally separate to complete the 3+1
dimensional super-multiplets.

The beta function of $\cal N=$ 4 SYM theory vanishes to all orders
in perturbation theory, $\beta=0$. This implies the theory is
conformal with conformal symmetry group $SO(4,2)$ also at the
quantum level. Moreover this theory has a global $SU(4)$ R~symmetry
group. The complete superconformal group is $SU(2,2| 4)$, of which
both $SO(4,2)$ and $SU(4)$ are bosonic subgroups.

\subsection{The AdS/CFT Correspondence }

The AdS/CFT correspondence which was first suggested by Maldacena
[12] in 1997, states that Type IIB string theory on $(AdS_5 ×
S^5)_N$ plus some appropriate boundary conditions (and possibly also
some boundary degrees of freedom) is dual to $\cal N = $4, d = 3 + 1
U(N) super-Yang-Mills. There are three different versions of this
conjecture [63], depending on the precise form of the limits taken.
In the strong version, Type IIB string theory on $AdS_5 \times
S^5$ is dual to $SU(N_c)$ SYM theory. The mild version relates
Classical type IIB strings on $AdS_5 \times
S^5$ to planar $SU(N_c)$ SYM theory.
But the mostly adopted form of the conjecture is the weak regime (in the
SUGRA limit) which specializes further to the case in which
$\lambda$ is large. In this limit, strongly coupled $\cal N = $4
$SU(N)$ Yang-Mills theory is mapped to supergravity on $AdS_5 \times
S^5$; the inverse string tension $\alpha'$ goes to zero.

A precise way in which the two theories can be mapped into each
other was proposed independently by Gubser, Klebanov and Polyakov
[41] and by Witten [56]. Since the boundary of the $AdS_5$ space,
namely $S^3 \times R$, is equivalent to $R^{3,1}$ , which is a copy
of the Minkowski space, plus a point at infinity, the authors
suggested a recipe to link the gravity theory in the bulk (AdS
space) to the field theory on the boundary (Minkowski space). In
this sense, the AdS/CFT correspondence can be considered as a
holographic projection of the supergravity theory in the bulk to the
field theory on the boundary.

Despite the fact that there is no proof of the AdS/CFT
correspondence taking account of its string-theoretical origin yet,
the huge amount of symmetry present almost guarantees that the
AdS/CFT correspondence should hold. When proceeding to less
symmetrical situations below, generalized gauge/gravity dualities
remain a conjecture though.

\subsection{QCD vs SYM }

It would be useful if the four dimensional theory on the boundary
were QCD, since this would allow us to explore its non-perturbative
regime by studying a perturbative dual theory. However, the field
theory described by the correspondence is a supersymmetric theory
with conformal invariance, while QCD has none of these features. The
most important differences between the two theories are[63]:

\begin{itemize}
\item QCD confines while SYM is not confining.
\item QCD has a chiral condensate while SYM has no chiral condensate.
\item QCD has a discrete spectrum while that of SYM is continuous.
\item QCD has a running coupling while SYM has a tunable coupling and is conformal.
\item QCD has quarks while SYM has adjoint matter.
\item QCD is not supersymmetric while SYM is maximally supersymmetric.
\item QCD has $N_c = 3$ in real life, while the AdS/CFT correspondence holds for large $N_c$.
\end{itemize}

However, the gauge/gravity duality can be expanded to more field
theories by changing the supergravity theory. This gives a
possibility to search for a field theory that is closer to QCD and
has a gravity dual.

\begin{itemize}
\item For example considering multiple D3-branes on curved backgrounds,
leads to an interesting family of $\cal N = $1 superconformal field
theories [64-65] which contain adjoint matter fields. Also, one can
introduce the confinement and broke the conformal symmetry by
deforming the background further. This leads to chiral symmetry
breaking and a running coupling constant [20].
\item Also, theories looking like $\cal N = $1 supersymmetric Yang-Mills
theory in the IR can be obtained by considering higher dimensional
D-branes wrapped on certain sub-manifolds of the ten dimensional
geometry [66-67].
\item Deformations of the geometry lead to non-supersymmetric,
non-conformal gauge theories which display confinement and chiral
symmetry breaking [18-19][68-71].
\item Fundamental matter can be added to the gauge theory by introducing
D7-branes [73]. In the quenched approximation, $N_f \ll N_c$, their
effect on the background geometry is ignored. Also, dynamical quarks
can be added to this geometry [73].
\item Recently, some phenomenologically models have been suggested which are motivated by the AdS/CFT but not within the full string theory
framework. These models are known as AdS/QCD [74-77].
\item Also, an approach similar to AdS/QCD is introduced based on the noncritical string theory in $d\ne 10$ dimensions [42,77-78].
\end{itemize}


\section{Holographic QCD from the non-critical string theory }

The key idea of construction of holographic models with flavors was
given by Karch and Katz [72]. In these models, two stacks of flavor branes, branes and anti-branes, are added to the geometry as a probe, so that the back reaction of the flavor branes is negligible (probe approximation). This approximation is reliable when $N_f \ll N_c $, where $N_c$ and $N_f$ refer to the
number of colors and flavors, respectively.

Of course, the brane/anti-brane system is unstable, since the branes and anti-branes will tend to annihilate. This is reflected in the presence of tachyons in the spectrum. But, it should make sense within the context of perturbation theory. The point where the tachyon field vanishes corresponds to a local maximum of the tachyon potential, and thus it is part of a classical solution. The one-loop effective action in an expansion around this solution should be well defined, even though the solution is unstable, and in particular it should have a well-defined phase. It was conjectured that at the minimum of the tachyon potential, the negative contribution to the energy density from the potential exactly cancels the sum of the tensions of the brane and the anti-brane, thereby giving a configuration of zero energy density (and hence restoring space-time supersymmetry). Therefore, the various gauge and gravitational anomalies, which arise as one-loop effects, cancel and as we expected  theory is perturbatively well-behaved [79-82].

In this section, we study a model which is similar in many aspects
to the SS model [18], a holographic model based on the critical
string theory. But, we try to solve some inconsistencies of the SS
model in describing the baryons via the non-critical $AdS_6$ model.

\subsection{$AdS_6$ model}

In the presented non-critical model, the gravity background is
generated by near-extremal $D4$ branes wrapped over a circle with
the anti-periodic boundary conditions. Two stacks of flavor branes,
namely $D4$ branes and anti-$D4$ branes, are added to this geometry
and are called flavor probe branes. The color branes extend along
the directions $t,x_1,x_2,x_3,$ and $\tau$ while the probe flavor
branes fill the whole Minkowski space and stretch along the radius
$U$ which is extended to infinity. The strings attaching a color
D4-brane to a flavor brane transform as quarks, while strings
hanging between a color $D4$ and a flavor $\overline{D4}$ transform
as anti-quarks. The chiral symmetry breaking is achieved by a
reconnection of the brane, anti-brane pairs. Under the quenched
approximation $(N_c \gg N_f)$, the reactions of flavor branes and
the color branes can be neglected. Just like the SS model, the
$\tau$ coordinate is wrapped on a circle and the anti-periodic
condition is considered for the fermions on the thermal circle. The
final low energy effective theory on the background is a
four-dimensional QCD-like effective theory with the global chiral
symmetry $U(N_f)_L\times U(N_f)_R$.

In this model, the near horizon gravity background at low energy is [45]
\begin{eqnarray}
ds^2=\left( \frac{U}{R} \right)^2 (-dt^2+dx_i dx_i+f(U) d\tau^2) +
\left( \frac{R}{U} \right)^2 \frac{dU^2}{f(U)},
\end{eqnarray}
where R is the radius of the AdS space. Also $f(U)$ and RR six-form
field strength, $F_{(6)}$ are defined by the following relations
\begin{eqnarray}
F_{(6)}&=&Q_c \left( \frac{U}{R} \right)^4 dt \wedge
dx_1\wedge dx_2 \wedge dx_3  \wedge du \wedge d\tau \,, \nonumber\\
f(U)&=&1-\left( \frac{U_{KK}}{U} \right)^5.
\end{eqnarray}
In order to obtain solutions of near extremal flavored $AdS_6$, the
values of dilaton and $R_{AdS}$ are considered as
\begin{eqnarray}
e^\phi &=& \frac{2}{3}\frac{Q_f}{Q_c^2}(\sqrt{1+\frac{6Q_c^2}{Q_f^2}}-1)\, ,\nonumber\\
R_{AdS}^2&=&\frac{90}{12+\frac{Q_f^2}{Q_c^2}-\frac{Q_f^2}{Q_c^2}\sqrt{1+\frac{6Q_c^2}{Q_f^2}}}.
\end{eqnarray}
This relation indicates that the $R_{AdS}$ and dilaton depend on the
ratio of the number of colors $(\sim Q_c)$ and flavors$(\sim Q_f)$.
Under the quenched approximation, the values of the dilaton and AdS
radius can be rewritten as,
\begin{eqnarray}
R_{AdS}^2=\frac{15}{2}\,\,\,\,,\,\,\,\,e^\phi &=
&\frac{2\sqrt2}{\sqrt3 Q_c},
\end{eqnarray}
where  $Q_c$ is proportional to the number of color branes, $N_c$.

To avoid singularity, the coordinate $\tau$ satisfies the following
periodic condition,
\begin{eqnarray}
\tau \sim \tau + \delta \tau\,\,,\qquad\qquad \delta\tau=\frac{4\pi
R^2}{5 U_{KK}}\,\,.
\end{eqnarray}
Also, the Kaluza-Klein mass scale of this compact dimension is
\begin{eqnarray}
M_{KK}=\frac{2\pi}{\delta\tau}= \frac{5}{2} \frac{ \ U_{KK}}{R^2},
\end{eqnarray}
and dual gauge field theory for this background is non
supersymmetric. Also, the Yang-Mills coupling constants can be
defined as a function of string theory parameters using the DBI
action as follows
\begin{eqnarray}
g_{YM}^2=\frac{g_s}{\mu_4 \,(2\pi \alpha')^2\, \delta \tau},
\end{eqnarray}
where $\alpha'=l_s^2$ is the Regge slope parameter and $l_s$ is the
string length. Also, the 't Hooft coupling is $\lambda=g_{YM}^2\,
N_c$.

\subsection{meson sector}

In AdS/QCD, there is a gauge field living in the bulk AdS whose
dynamics is dual to the meson sector of QCD such as pions and higher
resonances. The gauge field on the $D4$ brane includes five
components, $A_\mu (\mu=0,1,2,3)$ and $A_U$. The $D4$ brane action
is given by [49]
\begin{eqnarray}
S_{D4}= -\mu_4 \int d^5 x e^{-\phi} \sqrt{-\det(g_{MN}+2\pi\alpha'
F_{MN})}+S_{CS}, \label{FF}
\end{eqnarray}
where $F_{MN}=\partial_M A_N-\partial_N A_M-i[A_M,A_N],$
$(M,N=0,1,..5)$ is the field strength tensor, and the $A_M$ is the
$U(N_f)$ gauge field on the $D4$ brane. The second term in the above
action is the Chern-Simons action and $\mu_4=2\pi / (2\pi ls)^5$. It
is useful to define the new variable $z$ as
\begin{eqnarray}
U_z=(U_{KK}^5+U_{KK}^3\, z^2)^{1/5}.
\end{eqnarray}
Then by neglecting the higher order of $F^2$ in the expansion, the
$D4$ brane action can be written as [49]
\begin{eqnarray}
S_{D4}= -\widetilde \mu_4(2\pi\alpha')^2 \int d^4x dz\,[\,
\frac{R^4}{4U_z^{5/2}} \eta^{\mu\nu}\eta^{\rho\sigma} F_{\mu
\rho}F_{\nu \sigma}+\frac{25}{8}\frac{U_z^{9/2}}{U_{kk}^3}
\eta^{\mu\nu} F_{\mu z}F_{\nu z}\,] +\mathcal{O}(F^3) \, ,
\end{eqnarray}
where $\widetilde \mu_4$ is
\begin{eqnarray}
\widetilde \mu_4=\sqrt{\frac{3}{2}} \, \frac{N_c U_{KK}^{3/2}}{5
R^3}\, \mu_4.
\end{eqnarray}
The gauge fields $A_\mu$ ($\mu=0,1,2,3$) and $A_z$ have a mode
expansion in terms of complete sets $\{\psi_n(z)\}$ and
$\{\phi_n(z)\}$ as
\begin{eqnarray}
A_\mu(x^\mu,z)&=&\sum_{n} B_\mu^{(n)}(x^\mu) \psi_n(z) \ ,
\label{expand;Av}\\
A_z(x^\mu,z)&=&\sum_{n} \varphi^{(n)}(x^\mu) \phi_n(z) \ .
\label{expand;A}
\end{eqnarray}
After calculating the field strengths, the action (14) is rewritten
as
\begin{eqnarray}
S_{D4}&=&-\widetilde \mu_4 (2\pi\alpha')^2 \int d^4x dz \sum_{m,n}
\bigg[ \frac{R^4}{4U_z^{5/2}} F_{\mu\nu}^{(m)}F^{\mu\nu
(n)}\psi_m\psi_n +\nonumber \\ &&~~
\frac{25}{8}\frac{U_z^{9/2}}{U_{kk}^3} (
\partial_\mu \varphi^{(m)}\partial^\mu \varphi^{(n)}\phi_m\phi_n
~+B_\mu^{(m)}B^{\mu(n)}\dot\psi_m\dot\psi_n
-2\partial_\mu\varphi^{(m)}B^{\mu(n)}\phi_m\dot\psi_n )
\,\bigg],\nonumber\\
 \label{phiphiBB}
\end{eqnarray}
where the over dot denotes the derivative respect to the $z$
coordinate.

We introduce the following dimensionless parameters,
\begin{equation}
\widetilde z \equiv\frac{z}{U_{KK}} \ , ~~~ K(\widetilde z)\equiv
1+\widetilde z^2=\left(\frac{U_z}{U_{KK}}\right)^5 \ ,
\end{equation}
and find that the functions $\psi_n$ ($n\ge 1$) satisfy the
normalization condition as
\begin{equation}
\widetilde \mu_4 (2\pi\alpha')^2\frac{R^4}{U_{KK}^{3/2}} \int
d\widetilde z\,  K^{-1/2}\, \psi_n\psi_m =\delta_{nm} \ .
\end{equation}
Also, we suppose the functions $\psi_n$ ($n\ge 1$) satisfy the
following condition
\begin{eqnarray}
\widetilde \mu_4 (2\pi\alpha')^2\frac{R^4}{U_{KK}^{3/2}} \int
d\widetilde z \,K^{9/10} \,\partial_{\widetilde z}
\psi_m\,\partial_{\widetilde z} \psi_n =\lambda_n\delta_{nm} \, .
\end{eqnarray}
Using eqs. (20) and (21), an eigenvalue equation is obtained for the
functions $\psi_n$ ($n\ge 1$) as
\begin{equation}
-K^{1/2}\,\partial_{\widetilde z} \left(
K^{9/10}\,\partial_{\widetilde z} \psi_m\right)=\lambda_m\psi_m \ .
\end{equation}
The orthonormal condition for $\phi_n$ are as follows,
\begin{equation}
(\phi_m,\phi_n)\equiv \frac{25}{4}\,\widetilde \mu_4 (2\pi\alpha')^2
U_{KK}^{5/2} \int d\widetilde z \,K^{9/10}\,\phi_m \phi_n
=\delta_{mn} \,.
\end{equation}
We find that the functions $\phi_{(n)}$ and $\dot\psi_n $ are
related together. In fact, we can consider
$\phi_n=m_n^{-1}\,\dot\psi_n$ ($n\ge 1$). Also, there exists a
function $\phi_0=C/K^{9/10}$ which is orthogonal to $\dot\psi_n$ for
all $n\ge 1$
\begin{equation}
(\phi_0,\phi_n)\propto \int d\widetilde z\, \partial_{\widetilde
z}\psi_n =0 \ , ~~~~(\mbox{for}~n\ge 1)\,  . \label{phi0}
\end{equation}
We use the normalization condition $1=(\phi_0,\phi_0)$ to obtain the
normalization constant $C$. Finally by using an appropriate gauge
transformation, the action (14) becomes
\begin{eqnarray}
S_{D4}=- \int d^4x \, \bigg[\,
\frac{1}{2}\,\partial_\mu\varphi^{(0)}\partial^\mu\varphi^{(0)}
\sum_{n\ge 1}\bigg( \frac{1}{4} F_{\mu\nu}^{(n)}F^{\mu\nu (n)}
+\frac{1}{2} m_n^2\,B_\mu^{(n)}B^{\mu(n)} \bigg) \bigg], \label{S0}
\end{eqnarray}
where $B_\mu^{(n)}$ is a massive vector meson of mass $m_n\equiv
\lambda_n^{1/2} M_{KK}$ for all $n\ge 1$ and $\varphi^{(0)}$ is the
pion field, which is the Nambu-Goldstone boson associated with the
chiral symmetry breaking [49].

It is useful to make another gauge choice, namely the $A_z=0$ gauge.
Actually, we can transform to the new gauge through a syitable gauge
transformation and obtain the following new gauge fields,
\begin{eqnarray}
&&A_z(x^\mu,z)=0 \ ,\nonumber\\
&&A_\mu(x^\mu,z)=-\partial_\mu\varphi^{(0)}(x^\mu)\psi_0(z)
+\sum_{n\ge 1} B_\mu^{(n)}(x^\mu)
\psi_n(z) \ .\nonumber\\
\label{expAA}
\end{eqnarray}
Function $\psi_0(z)$ is calculated through
\begin{equation}
\psi_0(z)=\int^z_0 dz'\, \phi_0(z') =C\,U_{KK}\,\widetilde z
\,F_1(0.5,0.9,1.5,-\widetilde z^2),
\end{equation}
where $F_1$ is  well-known hypergeometric function. It should be
noted that the massless pseudo scalar meson appears in the
asymptotic behavior of $A_\mu$, since we have
\begin{eqnarray}
A_\mu(x^\mu,z)\rightarrow \pm 1.8 C U_{KK} \,
\partial_\mu\varphi^{(0)}(x^\mu)~~(\mbox{as}~ z\rightarrow\pm\infty).
\label{Abdry}
\end{eqnarray}
In order to calculate the meson spectrum, it is necessary to solve
the eq. (22) numerically by considering the normalization condition
(20).

Since eq. (22) is invariant under $\widetilde z\rightarrow
-\widetilde z$, we can assume $\psi_n$ to be an even or odd
function. In fact, the $B_{\mu}^{(n)}$ is a four-dimensional vector
and axial vector if $\psi_n$ is an even or odd function,
respectively. The Eq. (22) is solved numerically using the shooting
method to obtain the mass of lightest mesons. Our results are
compared with the results of the SS, KS, and DKS models and
experimental data in Table I. As is clear, our result are in good
agreement with the experimental data [49].

\begin{table}[htb]
\caption{\small . The ratio of the obtained eigenvalues of Eq. (22)
compared with the results of the KS [83], DKS [84], and SS model
[18] and the ratio of meson masses.} \center
\begin{tabular}{|c|c|c|c|c|c|}
\hline $\,\, k \,\,$ & $(\frac{\lambda_{k+1}}{\lambda_k})_{AdS_6}$ &
 $(\frac{\lambda_{k+1}}{\lambda_k})_{DKS}$&
 $(\frac{\lambda_{k+1}}{\lambda_k})_{KS}$ &
 $(\frac{\lambda_{k+1}}{\lambda_k})_{SS}$& $(\frac{\lambda_{k+1}}{\lambda_k})_{Exp}$
\\
\hline\hline 1 & 2.76 & 1.97 & 2.68 & 2.34 & ~2.51
\\
\hline 2 & 5.58 & 3.56 & 5.63 & 4.92 & ~3.65
\\
\hline 3 & 9.55 & 5.49 & 8.88 & 6.97 & ~4.45
\\
 \hline
\end{tabular}
\end{table}

\subsection{Pion effective action}

Now, we just consider the pion field in the gauge field expansion
and use the non-Abelian generalization of the DBI action to find the
effective pion action [49],
\begin{eqnarray}
S_{D4}= -\widetilde \mu_4 (2\pi\alpha')^2  \int d^4 x\,\mathop{\rm
tr}\nolimits \bigg( A(U^{-1}\partial_\mu U)^2
+B\,[U^{-1}\partial_\mu U,U^{-1}\partial_\nu U\,]^2 \bigg) \ ,
\label{pionEA}
\end{eqnarray}
where the coefficients A and B are defined by the following
relations [49]
\begin{eqnarray}
A&\equiv&2\, \frac{25}{8} \frac{1}{U_{KK}^3} \int d\widetilde z
\,U_z^{9/2} ( \partial_{\widetilde z} \widehat \psi_0(\widetilde
z))^2
=\frac{25 \, }{4 } \frac{U_{KK}^{1/2} }{3.6}\ ,\nonumber\\
B&\equiv&2\, \frac{R^4}{4} \int
dz\,\frac{1}{U_z^{5/2}}\psi_+^2(\psi_+-1)^2 =\frac{0.16 R^4
}{2U_{KK}^{3/2}} \ .
\end{eqnarray}
If we compare the Eq. (35) with the familiar action of the Skyrme
model [85]
\begin{eqnarray}
S=\int d^4 x\bigg(\frac{f_\pi^2}{4}\mathop{\rm tr}\nolimits (
 U^{-1}\partial_\mu U \,)^2+\frac{1}{32 e^2}
\mathop{\rm tr}\nolimits  [U^{-1}\partial_\mu U,U^{-1}\partial_\nu
U]^2 \bigg) \ , \label{Skyrme}
\end{eqnarray}
it is possible to calculate the pion decay constant $f_\pi$ and
dimensionless parameter $e$ in terms of the non-critical model
parameters [49]
\begin{eqnarray}
f_\pi^2= 4\,\widetilde \mu_4 (2\pi\alpha')^2 \,A=\sqrt{\frac32}\;
\frac{45\,\mu_4 (2\,\pi\,\alpha')^2}{3.6\,R^3} N_c M_{KK}^2\, ,
\label{fpi}
\end{eqnarray}
and
\begin{eqnarray}
\frac{1}{e^2}=32 \, \widetilde \mu_4 (2\pi\alpha')^2 \, B\, =
\sqrt{\frac38}\mu_4 (2\,\pi\,\alpha')^2\,R\, N_c\, . \label{e}
\end{eqnarray}
It is clear from the above equations that the parameters $f_\pi$ and
$e$ depend on $N_c$ as $f_\pi \sim \mathcal{O}(\sqrt{N_c})$ and
$e\sim\mathcal{O}(1/\sqrt{N_c})$ ,respectively. It is coincident
with the result obtained from the SS model and also QCD in large
$N_c$. We fix the $M_{KK}$ such that the $f_\pi\sim 93$ MeV for
$N_c=$3. So, we obtain $M_{KK}= 395$ MeV for our holographic model
[49]. It should be noted that $M_{KK}$ is the only mass scale of the
non-critical model below which the theory is effectively pure
Yang-Mills in four dimensions.

\subsection{Baryon in $AdS_6$ }

In this section we aim to introduce baryon configuration in the
non-critical holographic model. As is known, in the SS model the
baryon vertex is a $D4$ brane wrapped on a $S^4$ cycle. Here in
six-dimensional configuration, there is no compact $S^4$ sphere. So,
we introduce an unwrapped $D0$ brane as a baryon vertex instead
[86]. In analogy with the SS model, there is a Chern-Simons term on
the vertex world volume as
\begin{eqnarray}
S_{CS}\,\propto \, \int dt\, A_0(t), \label{e}
\end{eqnarray}
which induces $N_c$ units of electric charge on the unwrapped $D0$
brane. In accordance with the Gauss constraint, the net charge
should be zero. So, one needs to attach $N_c$ fundamental strings to
the $D0$ brane. In turn, the other side of the strings should end up
on the probe $D4$ branes. The baryon vertex looks like an object
with $N_c$ electric charge with respect to the gauge field on the
$D4$ brane whose charge is the baryon number. This $D0$ brane
dissolves into the $D4$ brane and becomes an instanton soliton [86].
It is important to know the size of the instanton in our model. In
the SS model, it is shown that the size of an instantonic baryon
goes to zero at large 't Hooft coupling limit which is one of the
problems of the SS model in describing the baryons [37].

Let us consider the DBI action in the Yang-Mills approximation for
the $D4$ brane
\begin{equation}
S_{YM}=-\frac14\;\mu_4  (2\pi\alpha')^2 \; \int
e^{-\phi}\,\sqrt{-g_{4+1}}  \;\rm tr \,F_{ mn}F^{ mn}\:.
\end{equation}
The induced metric on the $D4$ brane is
\begin{equation}
g_{4+1}=\left(\frac{U}{R}\right)^{2}\left(\eta_{\mu\nu}dx^{\mu}dx^{\nu}
+\left(\frac{R}{U}\right)^{4}\,\frac{dU^2}{f(U)}\right) \: .
\end{equation}
It is useful to define the new coordinate $w$
\begin{equation}
dw= \frac{R^{2}\,U^{1/2} \, dU }{\sqrt{U^5-U_{KK}^5}}\:.
\end{equation}
Using this coordinate, the metric (36) transforms to a conformally
flat metric
\begin{equation}
g_{4+1}=H(w)\left(dw^2+\eta_{\mu\nu}dx^{\mu}dx^{\nu}\right)\,\,,
\,\,\,\,H(w)=(\frac{U}{R})^2 \:.
\end{equation}
Also, the $w$ coordinate can be rewritten in terms of the $z$
coordinate introduced in Eq. (13) as
\begin{equation}
dw=\frac25 \frac{R^{2}\,U_{KK}^{3} \, dz }{(U_{KK}^5-U_{KK}^3\,
z^2)^{7/10}}\:.
\end{equation}
Note that in the new conformally flat metric, the fifth direction is
a finite interval $[-w_{max}, w_{max}]$ because
\begin{eqnarray}
w_{max}&=&\int_0^\infty \frac{R^{2}\,U^{1/2} \, dU
}{\sqrt{U^5-U_{KK}^5}} \simeq \frac{R^{2}}{U_{KK}}1.25 <\infty .
\end{eqnarray}
We can approximate $w$ near the origin $w\simeq 0$, as
\begin{equation}
w\simeq \frac{2}{5}\left(\frac{R}{U_{KK}}\right)^{2} z,
\end{equation}
and using relation (10), we obtain
\begin{equation}
w\simeq \frac{z}{M_{KK} \, U_{KK}} \,\,\,\,\,or\,\,\,\,\,M_{KK}\,
w\simeq \frac{z}{ U_{KK}},
\end{equation}
or equivalently,
\begin{equation}
U^5\simeq U_{KK}^5(1+M_{KK}^2\,w^2)\, .
\end{equation}
In analogy with the SS model, this relation implies that $M_{KK}$ is
the only mass scale that dictated the deviation of the metric from
the flat configuration and it is the only mass scale of the theory
in the low energy limit.(It should be noted that the $D4$ branes
come with two asymptotic regions at $w\rightarrow \pm w_{max}$
corresponding to the ultraviolet and infrared region near the $w
\simeq 0$.)

Equation (35) is rewritten in the conformally flat metric (38) as
\begin{eqnarray}
S_{YM}^{D4}&&=-\frac14\mu_4  (2\pi\alpha')^2  \int d^4x dw e^{-\phi} \left(\frac{U(w)}{R}\right)\rm tr F_{mn}F^{mn} \nonumber \\
&&=-\;\int dx^4 dw \;\frac{1}{4e^2(w)} \;\rm tr F_{mn}F^{mn} \: .
\end{eqnarray}
Thus, the position dependent electric coupling $e(w)$ of this five
dimensional Yang-Mills is equal to [30]
\begin{equation}
\frac{1}{e^2(w)}\equiv \frac{ \sqrt{3/2} \;
\mu_4\,(2\,\pi\,\alpha')^2\,R\,N_c}{5 }\; M_{KK} \left( \frac{U}{
U_{KK}}\right) \:.
\end{equation}
Also, for a unit instanton we have
\begin{equation}
\frac{1}{8\pi^2 }\int \rm tr F\wedge F=\frac{1}{16\pi^2} \int \rm tr
F_{mn} F^{mn} =1 .
\end{equation}
Inserting the above relations in the Eq. (44), we obtain the energy
of a point-like instanton localized at $w=0$ as
\begin{equation}
m_B^{(0)}= \frac{ \sqrt{3/2} \;4\pi^2 \mu_4\,(2\,\pi\,\alpha')^2 R
}{5}\;N_c  \; M_{KK}\:.
\end{equation}
By increasing the size of the instanton, more energy is needed
because $1/e^2(w)$ is an increasing function of $|w|$. So the
instanton tends to collapse to a point-like object. On the other
hand, $N_c$ fundamental strings attached to the $D4$ branes behave
as $N_c$ units of electric charge on the brane. The Coulomb
repulsions among them prefer a finite size for the instanton.
Therefore, there is a competition between the mass of the instanton
and Coulomb energy of fundamental strings. For a small instanton of
size $\rho$ with the density $D(x^i,w)\sim
\rho^4/(r^2+w^2+\rho^2)^4$, the Yang-Mills energy is approximated as
\begin{equation}\label{mass}
\sim \frac16\, m_B^{(0)}M_{KK}^2\rho^2\:,
\end{equation}
and the five dimensional Coulomb energy is
\begin{equation}\label{C}
\sim \frac12\times \frac{e(0)^2N_c^2}{10\pi^2\rho^2}\:.
\end{equation}
The size of a stable instanton is obtained by minimizing the total
energy [49]
\begin{equation}\label{size}
\rho^2_{baryon}\simeq \frac{1}{\sqrt{3/2} \;2\pi^2
\mu_4\,(2\,\pi\,\alpha')^2} \frac{1}{M_{KK}^2}\, .
\end{equation}
As it is stated in the previous section, in the SS model (the
critical version of dual QCD) the size of the instanton goes to zero
because of the large 't Hooft coupling limit. However in the
non-critical string theory, the 't Hooft coupling is of order one.
So, the size of the instanton is also of order 1 but it is still
smaller than the effective length of the fifth direction $\sim 1/
M_{KK}$ of the dual QCD.

\subsection{ Nucleon-Nucleon potential }

In the previous section, we demonstrated that the size of the baryon
in the non-critical holographic model is smaller than the scale of
the dual QCD and we can assume that the baryon is a point-like
object in five dimensions. Thus as a leading approximation, we can
treat it as a point-like quantum field in five dimensions. In the
rest of this paper, we will restrict ourselves to fermionic baryons
because we intend to study the nucleons. So, we consider odd
$N_c$ to study a fermionic spin $1/2$ baryon. We choose $N_c=3$ in
our numerical calculations for realistic QCD. Also, we will assume
$N_F=2$ and consider the lowest baryons which form the
proton-neutron doublet under $SU(N_F=2)$. All of these assumptions
lead us to introduce an isospin $1/2$ Dirac field, $\cal N$ for the
five dimensional baryon.

The leading 5D kinetic term for $\cal N$ is the standard Dirac
action in the curved background along with a position dependent mass
term for the baryon. Moreover, there is a coupling between the
baryon field and the gauge filed living on the flavor branes that
should be considered. Therefore, a complete action for the baryon
reads as [49]
\begin{eqnarray}
&&\int d^4 x dw\bigg[-i\bar{\cal N}\gamma^m D_m {\cal N} -i
m_b(w)\bar{\cal N}{\cal N} + g_5(w){\rho_{baryon}^2\over
e^2(w)}\bar{\cal N}\gamma^{mn}F_{mn}{\cal N} \bigg] -\nonumber \\
&&\int d^4 x dw {1\over 4 e^2(w)} \;\rm tr\, F_{mn}F^{mn}\,,
\label{5dfermion1}
\end{eqnarray}
where $D_m$ is a covariant derivative, $\rho_{baryon}$ is the size
of the stable instanton, and $g_5(w)$ is an unknown function with a
value at $w=0$ of  $2 \pi^2 /3$ [38]. $\gamma^m$ are the standard
$\gamma$  matrices in the flat space and $\gamma^{mn}=1/2
[\gamma^m,\gamma^n]$.

The factor $\rho_{baryon}^2\over e^2(w)$ is used for convenience.
Usually, the first two terms in the action are called the minimal
coupling and the last term in the first integral refers to the
magnetic coupling.

A four dimensional nucleon is the localized mode at $w\simeq 0$
which is the lowest eigenmode of a five dimensional baryon along the
$w$ direction. So, the action of the five dimensional baryon must be
reduced to four dimension. In order to do this, one should
perform the KK mode expansion for the baryon field $\mathcal
N(x_{\mu},w)$ and the gauge field $A(x_{\mu},w)$. The gauge field
has a KK mode expansion presented in Eqs. (16) and (17). The baryon
field also can be expanded as
\begin{eqnarray}
{\cal N}_{L,R}(x^\mu,w)=N_{L,R}(x^\mu)f_{L,R}(w), \label{5dfermion1}
\end{eqnarray}
where $N_{L,R}(x^\mu)$ is the chiral component of the four
dimensional nucleon field. Also the profile functions, $f_{L,R}(w)$
satisfy the following conditions:
\begin{eqnarray}
\partial_w f_L(w)+m_b(w) f_L(w) &=& m_B f_R(w)\:,\nonumber\\
-\partial_w f_R(w)+m_b(w) f_R(w) &=& m_B f_L(w)\:,
\end{eqnarray}
in the range $w\in[-w_{max},w_{max}]$, and the eigenvalue $m_B$ is
the mass of the nucleon mode, $N(x)$. Moreover, the eigenfunctions
$f_{L,R}(w)$ obey the following normalization condition
\begin{equation}
\int_{-w_{max}}^{w_{max}} dw\,\left|f_L(w)\right|^2 =
\int_{-w_{max}}^{w_{max}} dw\,\left|f_R(w)\right|^2 =1\:.
\end{equation}
It is more useful to consider the following second-order
differential equations for $f_{L,R}(w)$
\begin{eqnarray}
&&\left[-\partial^2_w -\partial_w m_b(w)+(m_b(w))^2\right]
f_L(w)=m_B^2 f_L(w)\:,\nonumber \\
&&\left[-\partial^2_w +\partial_w m_b(w)+(m_b(w))^2\right]
f_R(w)=m_B^2 f_R(w)\:. \label{eigeneq}
\end{eqnarray}

As we approach $w\to \pm w_{max}$, $m_b(w)$ diverges as $m_b(w) \sim
{1\over (w\mp w_{max})^2}$ and the above equations have normalizable
eigenfunctions with a discrete spectrum of $m_B$. Note that the term
$-\partial_w m_b(w)$ is asymmetric under $w\to -w$. It causes that
$f_L(w)$  tends to shift to the positive side of $w$ and the
opposite behavior happens for $f_R(w)$. It is important in the axial
coupling of the nucleon to the pions.

The gauge field  can be expanded in $A_z=0$ gauge as follows [49],
\begin{equation}
A_\mu(x,w)=i\alpha_\mu(x)\psi_0(w) +i\beta_\mu(x)+\sum_n
B_\mu^{(n)}(x)\psi_{(n)}(w)\:,
\end{equation}
where $\alpha_\mu$ and $\beta_\mu$ are related to the pion field
 $U(x)=e^{2i\pi(x)/f_\pi}$ by the following relations,
\begin{eqnarray}
&&\alpha_\mu(x)\equiv \{ U^{-1/2},\partial_{\mu} U^{1/2}\}\, ,\nonumber\\
&&\beta_\mu(x)\equiv \,\frac 12 [ U^{-1/2},\partial_{\mu} U^{1/2}]\,
.
\end{eqnarray}

Here, we use the above expansion along with the properties of
 $f_L(w)=\pm f_R(-w)$, $\psi_0$ and $\psi_{(n)}$ under the
$w\to-w$ transformation to calculate the four dimensional action. It
is worthwhile to note that again $\psi_{(2k+1)}(w)$ is even, while
$\psi_{(2k)}(w)$ is odd under $w\to-w$, corresponding to vector
$B_\mu^{(2k+1)}(x^\mu)$ and axial-vector mesons
$B_\mu^{(2k)}(x^\mu)$ respectively. For simplicity, we neglect the
Chern-Simons term in the baryon action, Eq. (51).

By inserting the mode expansion of the nucleon field and gauge field
into the baryon action [49],
\begin{eqnarray}
{\cal L}_{\rm Nucleon}= -i\bar N \gamma^\mu \partial_\mu N- im_B
\bar N N+ {\cal L}_{\rm vector}+{\cal L}_{\rm axial}\,,
\end{eqnarray}
where
\begin{eqnarray}
&&{\cal L}_{\rm vector}=-i\bar N \gamma^\mu \beta_\mu
 N-\sum_{k\ge 0}g_{V}^{(k)} \bar N \gamma^\mu  B_\mu^{(2k+1)} \,N
 +\sum_{k\ge 0}g_{dV}^{(k)} \bar N \gamma^{\mu \nu} \partial_\mu B_\nu^{(2k+1)}
 \,N\, ,
 \nonumber\\
&&{\cal L}_{\rm axial}=-\frac{i g_{A}}{2}\bar N  \gamma^\mu\gamma^5
\alpha_\mu N -\sum_{k\ge 1} g_{A}^{(k)} \bar N \gamma^\mu\gamma^5
B_\mu^{(2k)} N +\sum_{k\ge 0}g_{dA}^{(k)} \bar N \gamma^{\mu \nu}
\gamma^5
\partial_\mu B_\nu^{(2k)} \,N \, .\label{vector-axial-coupling}
\end{eqnarray}

Also, $g=g_{min}+g_{mag}$ stands for all the couplings. We neglect
the derivative couplings in the following calculations as a leading
approximation. The various minimal couplings constants
$g_{V,min}^{(k)},g_{A,min}^{(k)}$ as well as the pion-nucleon axial
coupling $g_{A,min}$ are calculated by the following suitable
overlap integrals of wave functions as
\begin{eqnarray}
g_{V,min}^{(k)}&=&\int_{-w_{max}}^{w_{max}}
dw\,\left|f_L(w)\right|^2
\psi_{(2k+1)}(w)\:,\nonumber\\
g_{A,min}^{(k)}&=&\int_{-w_{max}}^{w_{max}}
dw\,\left|f_L(w)\right|^2
\psi_{(2k)}(w)\:,\nonumber\\
g_{A,min}&=&2\int_{-w_{max}}^{w_{max}} dw\,\left|f_L(w)\right|^2
\psi_0(w)\:.
\end{eqnarray}

Also, we can compute the magnetic couplings using the following
integrals [49],
\begin{eqnarray}
g_{V,mag}^{(k)}&=& 2\,C_{mag} \int_{-w_{max}}^{w_{max}} dw \left(
\frac{g_5(w)}{g_5(0)}\right)\left( \frac{U(w)}{U_{KK}}\right)
\left|f_L(w)\right|^2\partial_w \psi_{(2k+1)}(w)\:, \nonumber\\
g_{A,mag}^{(k)}&=& 2\,C_{mag} \int_{-w_{max}}^{w_{max}} dw \left(
\frac{g_5(w)}{g_5(0)}\right)\left( \frac{U(w)}{U_{KK}}\right)
\left|f_L(w)\right|^2\partial_w \psi_{(2k)}(w)\:, \nonumber\\
g_{A,mag}&=& 4\, C_{mag}\int_{-w_{max}}^{w_{max}}  dw \left(
\frac{g_5(w)}{g_5(0)}\right)\left( \frac{U(w)}{U_{KK}}\right)
\left|f_L(w)\right|^2\partial_w \psi_0(w)\,,
\end{eqnarray}
where we define $C_{mag}$ as
\begin{eqnarray}
C_{mag}=\frac{ \sqrt{3/2} \mu_4\,(2\,\pi\,\alpha')^2}{5
}\,R\,N_c\;g_5(0)\, M_{KK} \;\rho_{baryon}^2\:.
\end{eqnarray}

Since the instanton carries only the non-Abelian field strength, the
iso-scalar mesons couple to the nucleon in a different formalism
than the iso-vector mesons. Therefore for the iso-scalar mesons,
such as the $\omega^{(k)}$ meson, only the minimal couplings
contribute
\begin{eqnarray}
g_{A}^{iso-scalar}&=&g_{A,min}\:,\nonumber\\
g_{A}^{(k),iso-scalar}&=&g_{A,min}^{(k)}\:,\nonumber\\
g_{V}^{(k),iso-scalar}&=&g_{V,min}^{(k)}\: .
\end{eqnarray}
However, the iso-vector mesons couple to the nucleon from both the
minimal and magnetic channels. Thus, iso-vector meson couplings are [49]
\begin{eqnarray}
g_{A}^{iso-vector}&=&g_{A,min}+g_{A,mag}\:,\nonumber\\
g_{A}^{(k),iso-vector}&=&g_{A,min}^{(k)}+g_{A,mag}^{(k)}\:,\nonumber\\
g_{V}^{(k),iso-vector}&=&g_{V,min}^{(k)}+g_{V,mag}^{(k)}\: .
\end{eqnarray}

The iso-scalar and iso-vector mesons have the same origin in the
five dimensional dynamics of the gauge field. In fact, if we write
the gauge field in the fundamental representation, we could
decompose the massive vector mesons as
\begin{equation}
B_\mu^{(2k+1)}=\left(\begin{array}{cc}1/2 &0 \\
0&1/2\end{array}\right) \omega^{(k)}_\mu+\rho_\mu^{(k)}\, ,
\end{equation}
where $\omega^{(k)}_\mu$ and $\rho_\mu^{(k)}$ are the iso-scalar and
the iso-vector parts of a vector meson, respectively. Since the
baryon is made out of $N_c$ product quark doublets, the above
composition for nucleon should be written as
\begin{equation}
B_\mu^{(2k+1)}=\left(\begin{array}{cc}N_c/2 &0 \\
0&N_c/2\end{array}\right) \omega^{(k)}_\mu+\rho_\mu^{(k)}\:.
\end{equation}
Therefore, there is an overall factor $N_c$ between the iso-scalar,
$\omega^{(k)}_\mu$ and iso-vector, $\rho_\mu^{(k)}$ mesons. Indeed,
there is a universal relation between the Yukawa couplings involving
the iso-scalar and iso-vector mesons
\begin{equation}
|g_{\omega^{(k)}NN}|\simeq  N_c \times |g_{\rho^{(k)}NN}|\,
.\label{prediction-gomega}
\end{equation}

We solve the eigenvalue Eq. (53) numerically using the shooting
method to obtain the wave function, $f_{L,R}$ and the mass, $m_B$ of
the nucleon. In order to do the numerical calculation, we assume
$N_c=3$ for realistic QCD. Also as was mentioned in the previous
section, we choose the value of $M_{KK}=0.395$ GeV to have the pion
decay constant $f_{\pi}=0.093$ GeV. We obtain the various couplings
by evaluating integrals (60) and (61) and compare some of our
results with the results of the SS model [37] in Table II.

\begin{table}[htb]
\caption{\small . Numerical results for axial and vector meson
couplings in the non-critical holographic model of QCD. The values
of vector couplings are compared with the SS model results[37]. }
\center
\begin{tabular}{|c|c|c|c|c|c|c|}
\hline $k$  &$g^{(k)}_{A,min}$ & $g_{A,mag}^{(k)}$ & $g^{(k),a}_{V,min}$ & $g^{(k),b}_{V,min}$ & $g_{V,mag}^{(k),a}$ & $g_{V,mag}^{(k),b}$ \\
 \hline\hline
0& 1.16 & 1.86 &8.30 & 5.933 & -1.988 & -0.816  \\
 \hline
1 &  1.07& 1.44 & 1.6488 &3.224& -6.83 & -1.988 \\
 \hline
2 &  0.96  & 0.862& 1.9 & 1.261& -7.44 &-1.932 \\
\hline
3  & 0.67 & 0.14& 0.688 & 0.311 & -4.60 & -0.969 \\
\hline
\end{tabular}
\\
(a) presented model results \\
(b) SS  model results
\end{table}

Also, using this non-critical model, the axial couplings are
obtained as
\begin{equation}
g_{A,mag}=1.582 \,\,\,,\,\,\, g_{A,min}\simeq 0 \, ,
\end{equation}
while in the previous analysis [18] using the SS model, these
couplings are reported as
\begin{equation}
g_{A,mag}=0.7\, \frac{N_c}{3} \,\,\,,\,\,\, g_{A,min}\simeq 0.13 \,
.
\end{equation}
If we choose $N_c=3$, then the SS model predicts $g_{A,mag}=0.7$ and
$g_{A}=0.83$. It should be noted  that the higher order of $1/{N_c}$
corrections can be used to improve this result but the lattice
calculations indicate that higher order of $1/{N_c}$ corrections are
suppressed. Our results are a good approximation of the experimental
data at leading order $g_{A}^{exp}=1.2670 \pm 0.0035$.

\subsubsection{ Nucleon-meson couplings }

Our holographic NN potential contains just the vector, axial-vector,
and pseudo-scalar meson exchange potentials which have the isospin
dependent and isospin independent components. The vector meson
($\omega^{(k)},\rho^{(k)}$), axial-vector meson ($f^{(k)},
a^{(k)}$), and pseudo-scalar meson ($\pi^{(k)},\eta'^{(k)} $)
couplings are related to the minimal and magnetic couplings as
follows
\begin{eqnarray}
g_{\omega^{(k)}NN}&\equiv & \frac{N_c\; g_{V}^{(k),iso-scalar}}{2} =\,\frac{N_c\; g_{V,min}^{(k)}}{2} \:,\\
g_{\rho^{(k)}NN}&\equiv  & \frac{ g_{V}^{(k),iso-vector}}{2}
=\,\frac{ g_{V,min}^{(k)}+g_{V,mag}^{(k)}}{2} \: ,\\
g_{f^{(k)}NN}&\equiv & \frac{N_c \, g_{A}^{(k),iso-scalar}}{2}
=\,\frac{N_c\, g_{A,min}^{(k)}}{2}, \\
g_{a^{(k)}NN}&\equiv & \frac{g_{A}^{(k),iso-vector}}{2} =\,\frac{
g_{A,min}^{(k)}+g_{A,mag}^{(k)}}{2}\:,\\
\frac{g_{\pi^{(k)}NN}}{2 m_N} M_{KK}  &\equiv & \frac{ g_{A}^{iso-vector}}{2 f_{\pi}} M_{KK} = \frac{ g_{Amin}+g_{A,mag} }{2  f_{\pi}}  M_{KK} ,\\
\frac{g_{\eta'^{(k)}NN}}{2 m_N} \,M_{KK} &\equiv  & \frac{N_c\,
g_{A}^{iso-scalar}}{2\, f_{\pi}}\,M_{KK} =\,\frac{N_c g_{A,min}}{2\,
f_{\pi}}\,M_{KK}.
\end{eqnarray}

All of the leading order meson-nucleon couplings are calculated
numerically and compared with the predictions of the four modern
phenomenological NN interaction models such as the AV 18 [8],
CD-Bonn [7], Nijmegen(93) [6] and Paris [5] potentials in Table III.
Also, results of the SS model are presented in this table. It is
necessary to mention here that the components of the
phenomenological models are very different in strength, and if
parameterized in terms of single meson exchange give rise to
effective meson-nucleon coupling strengths, which also are similar.
We explain different components of the NN potential below in detail.

The isospin dependent component of the vector potential which arises
from a $\rho$ meson exchange is roughly three times weaker than the
isospin independent component. In a chiral quark model, it is
expected to have $g_\omega=3\, g_\rho$, but the value of the
$\mathcal{R}=g_\omega/3\, g_\rho$ differs from the one in the above
phenomenological interaction models. It is 1.66 for the CD-Bonn, 1.5
for the Nijmegen, and 0.77 in the Paris model. This ratio is about
1.2 in the SS model and equals to $\mathcal{R}=1.33$ in our model.
Actually, the NN phase shifts uniformly require a larger
$\mathcal{R}$ than the chiral quark model prediction which is a
mystery. However in the resultant potential of the holographic QCD
model, it can be explained by the contribution of the magnetic
coupling in the vector channel.

\begin{table}[htb]
\caption{\small . The values of different effective meson-nucleon
couplings in the phenomenological interaction models [87], SS model
[18], and our model. } \center
\begin{tabular}{|c|c|c|c|c|c|c|}
\hline g &$V\,18$ & $CD-Bonn$ &  $Nijm\,(93)$  & $Paris$ & $SS\, model$ & $Our\, model$ \\
 \hline\hline
$g_{a^0} $ & 9.0 & 9.0 & 9.0 & 10.4 & - & - \\ \hline $g_\sigma$
&9.0 & 11.2 & 9.8 & 7.6 & - & - \\ \hline $g_\pi$    & 13.4 &
13.0&12.7&13.2 &  16.48   & 15.7  \\ \hline $g_\eta$   &  8.7 & 0.0
& 1.8 &11.7& 16.13 & 0.0 \\ \hline $g_\omega$ & 12.2 & 13.5& 11.7
&12.7 & 12.6 & 11.57 \\ \hline $g_\rho$   & -    &3.19 & 2.97
&- & 3.6 & 3.15 \\ \hline $g_{a^1}$& - & - &- &- & 3.94 & 1.51 \\
\hline
$g_{f^1}$  & - & - & - & - &  & 1.74 \\
\hline
\end{tabular}
\end{table}


\section{Holographic Light Nuclei}

In the holographic models, baryon is introduced as a D-brane wrapped
on a higher dimensional sphere in the curved space-time [17].
According to the fact that each nucleus is a set of A nucleons, so
the collection of the A baryon D-branes can describe a nucleus with
the mass number A. Then the dual gravity for the nucleus can be
obtained by applying the AdS/CFT correspondence. The U(A) gauge
theory living in the gravity dual of QCD is difficult to treat,
hence the large A limit is considered for this dual geometry which
corresponds to the heavy nuclei [88]. On the other hand, it is
necessary to use the nucleon-nucleon potential to study the
properties of light nuclei . In this section we aim to study the
holographic light nuclei such as $^{2}D$, $^{3}T$, $^{3}He$, and
$^{4}He$. For this purpose we consider a set of A instantonic
baryons as a nucleus. It is known that the nucleons are stabilized
at a certain distance in nuclei because of a binding force and a
strong repulsive force due to the light meson exchanges. We assume
that the nucleons have a uniform distribution in nucleus. Therefore
we consider a homogeneous distribution of D-branes in the
$\mathrm{R}^3$ space. In order to study the potential of nucleus, we
should regard the interaction between these D-branes. It was shown
that the size of baryon (instanton) is small and the interaction
between two instantons can be explained by the OBEP potential [49].
In this section we use this nucleon-nucleon potential to obtain the
potentials of light nuclei. Also we calculate the binding energy of
these nuclei. Then we impose different conditions on nucleon spins
in order to obtain some excited states of the $^{4}He$ nucleus.
Finally, we calculate the energy of these excited states and
estimate their excited energy.

\subsection{Nucleon-Nucleon Holography Potential}

Two particle scattering Phase shift in different partial waves as well as the bound state properties of deuteron are experimental data for
a two-nucleon system which identify the main properties of
nucleon-nucleon interaction. But the potentials attained
phenomenologically have many free parameters which are determined by
fitting to the experimental data. Various mesons and their
resonances play a special role in producing the nucleon-nucleon
potential with the following rules,
\begin{itemize}
  \item The long range part of the NN potential $( r > 3 fm)$ is mostly due to the one pion exchange machanism.
  \item Isoscalar mesons are responsible for the attractive interaction in the intermediate range of the potential $(0.7 < r < 2 fm)$.
  \item Exchanging the vector meson $\rho$ can explain the small attractive behavior of the odd-triplet state.
  \item Vector mesons produce the strong short range repulsion.
\end{itemize}
Then by considering these facts the general one boson exchange
nucleon-nucleon potential is written as [39],
\begin{equation}
V_{NN}=V_\pi+V_{\eta'}+\sum_{k=1}^\infty
V_{\rho^{(k)}}+\sum_{k=1}^\infty V_{\omega^{(k)}} +\sum_{k=1}^\infty
V_{a^{(k)}}+\sum_{k=1}^\infty V_{f^{(k)}},\label{18}
\end{equation}
which contains the pseudo-scalar$(\pi, \eta' )$, vector
$(\rho^{(k)}, \omega^{(k)} )$ and axial vector$(a^{(k)}, f^{(k)} )$
meson exchange potentials, respectively. It should be noted that
despite of the phenomenological NN interaction model, here we
compute all of the nucleon-meson couplings contributing in the above
potential using the noncritical holography model.

In our calculations, the leading parts of the potential come from the
pseudo scalar meson $\pi$, iso-scalar vector meson $\omega^{(k)}$,
iso-vector vector meson $\rho^{(k)}$ and iso-vector axial vector
meson $a^{(k)}$ exchange interactions,
\begin{eqnarray}
\frac{g_{\pi{\cal NN}}M_{KK}}{2m_{\cal N}} \sim g_{\omega^{(k)}\cal
NN} \sim \frac{\tilde g_{\rho^{(k)}\cal NN}M_{KK}}{2m_{\cal N}} \sim
g_{a^{(k)}\cal NN}.\label{19}
\end{eqnarray}

One pion exchange potential (OPEP) has the following form,
\begin{eqnarray}
V_{\pi}=\frac{1}{4\pi}\left(\frac{g_{\pi{\cal NN}}M_{KK}}{2m_{\cal
N}}\right)^2\frac{1}{M_{KK}^2r^3}S_{12}\vec\tau_1\cdot\vec\tau_2.\label{20}
\end{eqnarray}
Also, the holographic potentials for isospin singlet vector
mesons $\omega^{(k)}$, isospin triplet vector mesons $\rho^{(k)}$
and the triplet axial-vector mesons $a^{(k)}$ are,
\begin{eqnarray}
V_{\omega^{(k)}}= \frac{1}{4\pi}\; \left(g_{\omega^{(k)}\cal
NN}\right)^2\;m_{\omega^{(k)}}\; y_0(m_{\omega^{(k)}} r),\label{21}
\end{eqnarray}
\begin{eqnarray}
V_{\rho^{(k)}}\simeq
 \frac{1}{4\pi} \left(\frac{\tilde g_{\rho^{(k)}\cal
 NN}M_{KK}}{2m_{\cal N}}\right)^2 \frac{m_{\rho^{(k)}}^3}{3M_{KK}^2}\, [ 2y_0(m_{\rho^{(k)}} r)\vec\sigma_1\cdot\vec \sigma_2
-y_2(m_{\rho^{(k)}} r) S_{12}(\hat {r})]\vec\tau_1\cdot\vec\tau_2
,\label{22}
\end{eqnarray}
and
\begin{eqnarray}
V_{a^{(k)}}\simeq \frac{1}{4\pi}\,\left({g_{a^{(k)}\cal
NN}}\right)^2 \frac{m_{a^{(k)}}}{3} \times [ -2y_0(m_{a^{(k)}}
r)\vec\sigma_1\cdot\vec \sigma_2 +y_2(m_{a^{(k)}} r) S_{12}(\hat
{r})]\vec\tau_1\cdot\vec\tau_2  .\label{23}
\end{eqnarray}

In the above equations we have,
\begin{eqnarray}
S_{12}=3(\vec\sigma_1\cdot\hat r)(\sigma_2\cdot\hat r)
-\vec\sigma_1\cdot\vec\sigma_2,\label{24}
\end{eqnarray}
and
\begin{eqnarray}
y_0(x)=\frac{e^{-x}}{x},~~~~~~~
y_2(x)=\left(1+\frac{3}{x}+\frac{3}{x^2}\right)\frac{e^{-x}}{x}\,
.\label{25}
\end{eqnarray}
The masses of all mesons are of the order $M_{KK}$ and
$m_{\rho^{(k)}}=m_{\omega^{(k)} } < m_{a^{(k)}}$. Also, the mass of
pion in the holographic model is zero and its coupling constant to
the nucleon in our approach is 15.7.

Finally, the holographic nucleon-nucleon potential becomes [51-53],
\begin{equation}
V^{holography}_{NN}=V_C(r)+(V_T^{\sigma}(r)
\vec\sigma_1\cdot\vec\sigma_2+ V_T^{S}(r)
S_{12})\,\vec\tau_1\cdot\vec\tau_2 .\label{26}
\end{equation}
where
\begin{equation}
V_C(r) =\sum_{k=1}^{P} \frac{1}{4\pi}\; \left(g_{\omega^{(k)}\cal
NN}\right)^2\;m_{\omega^{(k)}}\; y_0(m_{\omega^{(k)}} r)m,
\label{27}
\end{equation}
\begin{eqnarray}
V_T^{\sigma}(r)&&=\sum_{k=1}^{P} \frac{1}{4\pi} \left(\frac{\tilde
g_{\rho^{(k)}\cal NN}M_{KK}}{2m_{\cal N}}\right)^2
\frac{m_{\rho^{(k)}}^3}{3M_{KK}^2} [ 2y_0(m_{\rho^{(k)}}
r)]\nonumber \\ && +\sum_{k=1}^{P}
\frac{1}{4\pi}\,\left({g_{a^{(k)}\cal NN}}\right)^2
\;\frac{m_{a^{(k)}}}{3} \;[ -2y_0(m_{a^{(k)}} r)],\label{28}
\end{eqnarray}
and,
\begin{eqnarray}
V_T^{S}(r)&&=\frac{1}{4\pi}\left(\frac{g_{\pi{\cal
NN}}M_{KK}}{2m_{\cal
N}}\right)^2\frac{1}{M_{KK}^2r^3}\;\nonumber\\
&& +\sum_{k=1}^{P} \frac{1}{4\pi} \left(\frac{\tilde
g_{\rho^{(k)}\cal NN}M_{KK}}{2m_{\cal N}}\right)^2
\frac{m_{\rho^{(k)}}^3}{3M_{KK}^2} [ -y_2(m_{\rho^{(k)}} r)] \; \nonumber\\
&&+\sum_{k=1}^{P} \frac{1}{4\pi}\,\left({g_{a^{(k)}\cal
NN}}\right)^2 \;\frac{m_{a^{(k)}}}{3} \;[ y_2(m_{a^{(k)}} r)]
.\label{29}
\end{eqnarray}
It is shown that in the SS model,at the large enough distances,
$p\simeq \sqrt{\lambda/10}$ is an acceptable value for these
potentials. We consider the ten first terms of the above potentials
in our numerical calculations both in SS and $AdS_6$ models.

In order to calculate the NN potential, the nucleon-meson coupling
constants are needed. These couplings are calculated using the SS
model at the large $\lambda\, \Nc$ limit and presented in Table IV.

\begin{table}[htb]
\caption{\small . The values of meson-nucleon couplings and mass of
mesons in the SS model. The values of $\Nc=3$, $\lambda=400$ and
$m_N=550 MeV$ are supposed in calculations. } \center
\begin{tabular}{|c|ccccc|}
\hline
k &   $m_{\omega^k}$ & $m_{a^k}$ &  $g_{\omega^k}$  & $g_{\rho^k}$ & $g_{a^k}$  \\
 \hline \hline
0 &  0.818  & 1.25 &   2.1165&  0.7055 & 0.8140 \\

1  & 1.69 &   2.13& 1.9312& 0.6437 & 1.4202 \\
2  & 2.57 &   3.00  & 1.8888 & 0.6296& 2.0178 \\
3& 3.44& 3.87 & 1.8740 & 0.6246 & 2.6067\\
4 &  4.30  & 4.73 &1.8680& 0.6226& 3.1956 \\
5 &5.17& 5.59& 1.8636& 0.6212& 3.7931 \\
6 &6.03& 6.46 &1.8619& 0.6206 &4.3734 \\
7& 6.89& 7.32 &1.8602& 0.6200& 4.9623 \\
8 &7.75   & 8.19& 1.8602& 0.6200 &5.5512 \\
9 &8.62 &9.05& 1.8593& 0.6197& 6.1401
\\ \hline
\end{tabular}
\end{table}

Also, we calculate the coupling values in the noncritical $AdS_6$
background. The obtained results are presented in Tables V and IV .
In the follow, we calculate the light nuclei potentials using the NN
holography potentials coming from both SS and $AdS_6$ models.

\begin{table}[htb]
\caption{\small . Numerical results of vector meson couplings to the
nucleon for ten lowest mesons using the $AdS_6$ model. Meson masses
are in the $\Mkk$ unit. } \center
\begin{tabular}{|c|ccccc|}
\hline
k &   $g_{V,mag}^k$ & $g_{V,min}^k$ &  $g_{\omega^k}$  & $g_{\rho^k}$ & $m_{2k+1}^2$  \\
 \hline \hline
0  & -1.9889& 7.7251 & 11.5727 &2.8630& 0.5516 \\
  1  & -6.8384 &7.3315&
10.9974 &0.24 &   3.0593 \\
  2 &  -7.4493& 7.2420&  10.863&  0.1036  &7.6012 \\
 3&
-4.6067& 7.2211 & 10.8317& 1.3072&  14.1905  \\
4 &  -4.4327& 7.2147&  10.8222&
1.3910&  22.8274 \\
  5 &  -6.6083& 7.2133&  0.8200&  0.3024&  33.5191 \\
 6&
-6.1778& 7.2137 & 10.8206& 0.5179 & 46.2717 \\
  7 &  -4.0509 &7.1740 & 10.7611&
1.5616 & 60.3053 \\
  8  & -4.4701 &7.1725&  10.7589 &1.3512 & 76.8821 \\
  9&
-6.5703 &7.1714 & 10.7572 &0.3005 & 95.4673
\\ \hline
\end{tabular}
\end{table}

\begin{table}[htb]
\caption{\small Numerical results of axial-vector meson couplings to
the nucleon for ten lowest mesons using the $AdS_6$ model. Meson
masses are in the $\Mkk$ unit.} \center
\begin{tabular}{|c|ccccc|}
\hline
k &   $g_{A,mag}^k$ & $g_{A,min}^k$ &  $g_{a^k}$  & $g_{f^k}$ & $m_{2k}^2$  \\
 \hline \hline
0  & 4.2648 & 1.1659 & 2.7154&  1.7489 & 1.5389 \\
   1 &  5.3813 & 1.0718&
3.2301 & 1.6189&  5.0877  \\
  2  & 7.8574 & 0.9692&  4.4133 & 1.4539 & 10.6404  \\
  3 &
10.3344& 0.6713 & 5.5028 & 1.0069 & 18.2525  \\
  4  & 12.8068 & 0.4188 & 6.6128&
0.6282 & 27.9160  \\
 5 &  15.2780 & 0.3020 & 7.7900 & 0.4531 & 39.6300  \\
  6 &
17.7493  & 0.2743 & 9.0118 & 0.4115 & 53.4224  \\
  7  & 20.0849 & 0.2620 & 10.1734&
0.3930 & 68.3462  \\
  8 &  22.528 & 0.2359 & 11.3820 & 0.3539 & 85.9293  \\
  9 &
24.9705 & 0.2061 & 12.5885& 0.3092 & 105.5220
\\ \hline
\end{tabular}
\end{table}

\subsection{Holographic Deuteron}
Deuteron is the only bound state of two-nucleons system with the
isospin $T=0$, total spin $S=1$, spin-parity $1^+$, and binding
energy $E_B=2.225\,MeV$. In our holographic model, we suppose that
deuteron is made of two instantonic baryons with $\Nf=2$ and $\Nc=3$
which are located at relative distance r in the $R^3$ space and
consider the following potential for the deuteron
\begin{equation}
V^{holography}_{deuteron}=V_C+(V_T^{\sigma}\vec\sigma_1\cdot\vec\sigma_2+
V_T^{S} S_{12})\,\vec\tau_1\cdot\vec\tau_2 .\label{25}
\end{equation}
where $V_C(r), V_T^{\sigma}(r)$ and $V_T^{S}(r)$ are presented in
equations (85), (86), (87) ,respectively. The super selection rules
propose that
\begin{equation}
S_{12}=2,\,\,\,\,\,\,\,\vec\sigma_1\cdot\vec\sigma_2 =
1,\,\,\,\,\,\,\, \vec\tau_1\cdot\vec\tau_2=-3.\label{29}
\end{equation}

The deuteron potential is calculated numerically using the results
of the both SS model and $AdS_6$ model. The minimum of this
potential is considered as the deuteron binding energy. We choose
the $N_c=3,\, \lambda=400$ and $m_N=550\, MeV$ in the SS model.

As we know, the t'Hooft parameter is of order one in noncritical
holographic models. So, we choose the  $N_c=3,\, \lambda=1$ values
in our calculations in the $AdS_6$ model. Also, we use the obtained
value for the nucleon mass $m_N=920\, MeV$ in this model which is
very close to the real value of nucleon mass. Numerical results are
shown in Table VII.

\subsection{Holographic Tritium}

The next nucleus we considered here, is tritium which is composed of
three nucleons, two neutrons and one proton. We propose a
equilateral triangular configuration for the tritium  nucleus in
which the distance between each two nucleons is r. We suppose that
the total potential of the nucleus is the sum of the all
nucleon-nucleon interaction potentials which are parameterized in
terms of a single parameter r. In fact, the radius of nucleus can be
expresses in terms of parameter r. Finally, we write the following
potential for the tritium,
\begin{eqnarray}
V^{holography}_{Tritium}&=& V_{12}+ V_{13}+ V_{23} \nonumber\\
&=&3\,V_C(r)+(V_T^{\sigma}(r)\vec\sigma_1\cdot\vec\sigma_2+
V_T^{S}(r)
S_{12})\,\vec\tau_1\cdot\vec\tau_2 \nonumber\\
&+&(V_T^{\sigma}(r)\vec\sigma_1\cdot\vec\sigma_3+ V_T^{S}(r)
S_{13})\,\vec\tau_1\cdot\vec\tau_3 \nonumber\\
&+&(V_T^{\sigma}(r)\vec\sigma_2\cdot\vec\sigma_3+ V_T^{S}(r)
S_{23})\,\vec\tau_2\cdot\vec\tau_3.\label{30}
\end{eqnarray}
The super selection rules for this three-nucleon systems imply that
\begin{eqnarray}
&&S_{12}=2\,\,,\,\,\,\,\,\vec\sigma_1\cdot\vec\sigma_2 =1
\,\,,\,\,\,\,\,\,\,\,\,\, \vec\tau_1\cdot\vec\tau_2=-3 \nonumber\\
&&S_{13}=0\,\,,\,\,\,\,\,\vec\sigma_1\cdot\vec\sigma_3 =-3
\,\,,\,\,\,\,\, \vec\tau_1\cdot\vec\tau_3= -3\nonumber\\
&&S_{23}=0\,\,,\,\,\,\,\,\vec\sigma_2\cdot\vec\sigma_3 =-3
\,,\,\,\,\,\,\, \vec\tau_2\cdot\vec\tau_3=1 \, \, .\label{31}
\end{eqnarray}

\subsection{Holographic $^3He$}

In order to study the $^{3}He$ nucleus, it is necessary to add the
repulsive Coulomb energy to the potential. So, we consider the
following potential for the $^{3}He$ nucleus,
\begin{eqnarray}
V^{holography}_{^{3}He}&=& V_{12}+ V_{13}+ V_{23} \nonumber\\
&=&3 V_C (r)+E_c(r)\nonumber\\
&+&(V_T^{\sigma}(r)\vec\sigma_1\cdot\vec\sigma_2+ V_T^{S}(r)
S_{12})\,\vec\tau_1\cdot\vec\tau_2 \nonumber\\
&+&(V_T^{\sigma}(r)\vec\sigma_1\cdot\vec\sigma_3+ V_T^{S}(r)
S_{13})\,\vec\tau_1\cdot\vec\tau_3 \nonumber\\
&+&(V_T^{\sigma}(r)\vec\sigma_2\cdot\vec\sigma_3+ V_T^{S}(r)
S_{23})\,\vec\tau_2\cdot\vec\tau_3,\label{31}
\end{eqnarray}
where $E_c(r)$ is the Coulomb repulsion between two instantons
carrying $N_c$ unit of electric charge [14]. The protons of $^{3}He$
in the ground state have the opposite spin directions, so the
spin-parity of $^{3}He$ nucleus in the ground state is
$\frac{1}{2}^+$. On the other hand, we should have $L+S+T=1$ for a
system of two nucleons. It is well known that the nucleons in the
ground state of the $^{3}He$ are in $L=0$ state. So, by using the
super selection rules we obtain,
\begin{eqnarray}
&&S_{12}=0  \,\,,\,\,\,\,\,\vec\sigma_1\cdot\vec\sigma_2 =-3
\,\,,\,\,\,\,\, \vec\tau_1\cdot\vec\tau_2=1
\nonumber\\
&&S_{13}=2  \,\,,\,\,\,\,\,\vec\sigma_1\cdot\vec\sigma_3 =1 \,\,\,\,
\,\,,\,\,\,\,\,\, \vec\tau_1\cdot\vec\tau_3= -3
\nonumber\\
&&S_{23}=0  \,\,,\,\,\,\,\,\vec\sigma_2\cdot\vec\sigma_3 =-3
\,,\,\,\,\,\,\, \vec\tau_2\cdot\vec\tau_3=1 \, \, .\label{33}
\end{eqnarray}

If we consider another sets of nucleons in $^{3}He$ such that the
spin of protons be in a parallel direction, the spin- parity of
$^{3}He$ nucleus should be equal to $( \frac{3}{2})^+$. By super
selection rules, we have
\begin{eqnarray}
&&S_{12}=2  \,\,,\,\,\,\,\,\vec\sigma_1\cdot\vec\sigma_2 =1
\,\,,\,\,\,\,\, \vec\tau_1\cdot\vec\tau_2=1
\nonumber\\
&&S_{13}=2  \,\,,\,\,\,\,\,\vec\sigma_1\cdot\vec\sigma_3= 1
\,\,,\,\,\,\,\, \vec\tau_1\cdot\vec\tau_3=-3
\nonumber\\
&&S_{23}=2\,\,,\,\,\,\,\,\vec\sigma_2\cdot\vec\sigma_3 =1
\,,\,\,\,\,\,\, \vec\tau_2\cdot\vec\tau_3=-3 \, \, .\label{34}
\end{eqnarray}
We found that there is no bound state in this case both in SS and
$AdS_6$ models. Thus we conclude that there is no excited state for
the $^{3}He$ nucleus.

\subsection{Holographic $^4He$}

There are more than one possible configuration for a system with
four nucleons. The most symmetric configurations are tetrahedron,
diamond, and square configurations. If we suppose that the nucleons
are located in the corners of a tetrahedron configuration which is
made of four equilateral triangles, the distance between any two
nucleons is similar. So, the total potential is sum of the 6
nucleon-nucleon interactions with a same relative distance. But, we
know that the Coulomb interaction between protons prefers a larger
proton-proton distance than neutron-neutron or neutron-proton
distances. If two protons sit on the contrary corners of a square,
then the proton-proton distance is larger than the neutron-proton
distance. So, we consider the square configuration for the $^{4}He$
nucleus and write the potential of $^{4}He$ nucleus as the following
form
\begin{eqnarray}
V^{holography}_{^{4}He}&=& V_{12}+ V_{13}+ V_{14}+ V_{23}+ V_{24}+ V_{34}\nonumber\\
&=& 4 V_C (r)+ 2 V_C (\sqrt{3}r)+ E_c(\sqrt{2}r)\nonumber\\
&+&(V_T^{\sigma}(r)\vec\sigma_1\cdot\vec\sigma_2+
V_T^{S}(r)S_{12})\,\vec\tau_1\cdot\vec\tau_2
\nonumber\\
&+&(V_T^{\sigma}(r)\vec\sigma_1\cdot\vec\sigma_3+
V_T^{S}(r)S_{13})\,\vec\tau_1\cdot\vec\tau_3
\nonumber\\
&+&(V_T^{\sigma}(\sqrt{2}r)\vec\sigma_1\cdot\vec\sigma_4+ V_T^{S}(\sqrt{2}r)S_{14})\,\vec\tau_1\cdot\vec\tau_4 \nonumber\\
&+&(V_T^{\sigma}(\sqrt{2}r)\vec\sigma_2\cdot\vec\sigma_3+
V_T^{S}(\sqrt{2}r)S_{23})\,\vec\tau_2\cdot\vec\tau_3
\nonumber\\
&+&(V_T^{\sigma}(r)\vec\sigma_2\cdot\vec\sigma_4+
V_T^{S}(r)S_{24})\,\vec\tau_2\cdot\vec\tau_4
\nonumber\\
&+&(V_T^{\sigma}(r)\vec\sigma_3\cdot\vec\sigma_4+
V_T^{S}(r)S_{34})\,\vec\tau_3\cdot\vec\tau_4 .\label{35}
\end{eqnarray}

\subsubsection{Ground State}

It is well known from the Pauli exclusion rule that the spins of two
protons (neutrons) have opposite directions and the $^{4}He$ nucleus
in the ground state has the spin-parity $0^+$. The super selection
rules for this structure imply that,
\begin{eqnarray}
&& S_{12}=0   \,\,,\,\,\,\,\,  \vec\sigma_1\cdot\vec\sigma_2 =-3
\,\,,\,\,\,\,\,  \vec\tau_1\cdot\vec\tau_2=1
\nonumber\\
&& S_{13}= 2  \,\,,\,\,\,\,\,  \vec\sigma_1\cdot\vec\sigma_3 =1
\,\,\,\,\,\,,\,\,\,\,\,  \vec\tau_1\cdot\vec\tau_3=-3
\nonumber\\
&& S_{14}=0  \,\,,\,\,\,\,\,  \vec\sigma_1\cdot\vec\sigma_4 = -3
\,\,,\,\,\,\,\,  \vec\tau_1\cdot\vec\tau_4=1
\nonumber\\
&& S_{23}=0   \,\,,\,\,\,\,\,  \vec\sigma_2\cdot\vec\sigma_3 =-3
\,\,,\,\,\,\,\,  \vec\tau_2\cdot\vec\tau_3=1
\nonumber\\
&& S_{24}= 2  \,\,,\,\,\,\,\,  \vec\sigma_2\cdot\vec\sigma_4 = 1
\,\,\,\,\,\,,\,\,\,\,\,  \vec\tau_2\cdot\vec\tau_4=-3
\nonumber\\
&& S_{34}=0   \,\,,\,\,\,\,\,  \vec\sigma_3\cdot\vec\sigma_4 = -3
\,\,,\,\,\,\,\,  \vec\tau_3\cdot\vec\tau_4=1 .\label{36}
\end{eqnarray}

\subsubsection{Excited States}

Also, the potential of $^{4}He$ is obtained for its excited states
with $(2^-,T=1)$, $(2^-,T=0)$ and  $(1^-,T=1)$  by considering
various structures for the spin-parity of nucleons. The holographic
potential for each excited state has a minimum. The excited energies
of these states can be regarded as the difference between the
minimum point of potential in each state and the binding energy of
nucleus.

If two nucleons ( two protons or neutron ) have the same spin
directions and occupy the level $L=1$, we find the excited level
with $2^-$, $T=1$ and excited energy $E_{ex}=23.330 \, MeV$. Super
selection rules for this state lead to,
\begin{eqnarray}
&& S_{12}=2   \,\,,\,\,\,\,\,  \vec\sigma_1\cdot\vec\sigma_2 =1
\,\,,\,\,\,\,\,  \vec\tau_1\cdot\vec\tau_2=1
\nonumber\\
&& S_{13}=0   \,\,,\,\,\,\,\,  \vec\sigma_1\cdot\vec\sigma_3 =-3
\,\,,\,\,\,\,\,  \vec\tau_1\cdot\vec\tau_3=-3
\nonumber\\
&& S_{14}=2   \,\,,\,\,\,\,\,  \vec\sigma_1\cdot\vec\sigma_4 = 1
\,\,,\,\,\,\,\,  \vec\tau_1\cdot\vec\tau_4=1
\nonumber\\
&& S_{23}=0   \,\,,\,\,\,\,\,  \vec\sigma_2\cdot\vec\sigma_3 =-3
\,,\,\,\,\,\,\,  \vec\tau_2\cdot\vec\tau_3=1
\nonumber\\
&& S_{24}=2   \,\,,\,\,\,\,\,  \vec\sigma_2\cdot\vec\sigma_4 = 1
\,\,,\,\,\,\,\,  \vec\tau_2\cdot\vec\tau_4=1
\nonumber\\
&& S_{34}=0   \,\,,\,\,\,\,\,  \vec\sigma_3\cdot\vec\sigma_4 =-3
\,\,,\,\,\,\,\,  \vec\tau_3\cdot\vec\tau_4=-3 .\label{38}
\end{eqnarray}
Numerical values for the potential of this excited state are shown
in Table . For this state we obtain $E_{Excited}=25.1005\,MeV$ using
the value $M_{KK}=395\,MeV$. While such excited state is not
predicted by the SS model [52].

In another structure, we suppose that the spins of two protons ( or
neutrons) have the same directions and one of them occupies the
$L=1$ level. In this case, the spin-parity of the state is $2^-$. It
may be treated as excited state of $^{4}He$ nucleus with spin-parity
and isospin $2^-$, $T=0$ and the excited energy $E_{ex}=21.840 \,
MeV$. In order to calculate its holographic potential, following
values which are obtained from the super selection rules have been
used
\begin{eqnarray}
&& S_{12}=2   \,\,,\,\,\,\,\,  \vec\sigma_1\cdot\vec\sigma_2 =1
\,\,,\,\,\,\,\,  \vec\tau_1\cdot\vec\tau_2=-3
\nonumber\\
&& S_{13}=0   \,\,,\,\,\,\,\,  \vec\sigma_1\cdot\vec\sigma_3 = -3
\,\,,\,\,\,\,\,  \vec\tau_1\cdot\vec\tau_3=1
\nonumber\\
&& S_{14}=2   \,\,,\,\,\,\,\,  \vec\sigma_1\cdot\vec\sigma_4 =1
\,\,,\,\,\,\,\,  \vec\tau_1\cdot\vec\tau_4=1
\nonumber\\
&& S_{23}=0   \,\,,\,\,\,\,\,  \vec\sigma_2\cdot\vec\sigma_3 =-3
\,,\,\,\,\,\,\,  \vec\tau_2\cdot\vec\tau_3=1
\nonumber\\
&& S_{24}=2   \,\,,\,\,\,\,\,  \vec\sigma_2\cdot\vec\sigma_4 =1
\,\,,\,\,\,\,\,  \vec\tau_2\cdot\vec\tau_4=1
\nonumber\\
&& S_{34}=0   \,\,,\,\,\,\,\,  \vec\sigma_3\cdot\vec\sigma_4 =-3
\,\,,\,\,\,\,\,  \vec\tau_3\cdot\vec\tau_4=-3 .\label{37}
\end{eqnarray}

The exited energy for this state is obtained about
$E_{excited}=21.8237\, MeV$  using the value $M_{KK}=395\,MeV$.

If the spin of proton ( neutron ) in the $L=1$ level couples with
the spin of the proton ( neutron ) in the $L=0$ state, we find
another excited state with the $1^-$, $T=1$ and the measured excited
energy $E_{ex}=23.640 \, MeV$. In this case we have,
\begin{eqnarray}
&& S_{12}= 2  \,\,,\,\,\,\,\,  \vec\sigma_1\cdot\vec\sigma_2 =1
\,\,,\,\,\,\,\,  \vec\tau_1\cdot\vec\tau_2=1
\nonumber\\
&& S_{13}= 0  \,\,,\,\,\,\,\,  \vec\sigma_1\cdot\vec\sigma_3 = -3
\,\,,\,\,\,\,\,  \vec\tau_1\cdot\vec\tau_3=-3
\nonumber\\
&& S_{14}=0   \,\,,\,\,\,\,\,  \vec\sigma_1\cdot\vec\sigma_4 = -3
\,\,,\,\,\,\,\,  \vec\tau_1\cdot\vec\tau_4=-3
\nonumber\\
&& S_{23}= 0  \,\,,\,\,\,\,\,  \vec\sigma_2\cdot\vec\sigma_3 = -3
\,,\,\,\,\,\,\,  \vec\tau_2\cdot\vec\tau_3=1
\nonumber\\
&& S_{24}=0   \,\,,\,\,\,\,\,  \vec\sigma_2\cdot\vec\sigma_4 =-3
\,\,,\,\,\,\,\,  \vec\tau_2\cdot\vec\tau_4=1
\nonumber\\
&& S_{34}= 2  \,\,,\,\,\,\,\,  \vec\sigma_3\cdot\vec\sigma_4 =1
\,\,,\,\,\,\,\,  \vec\tau_3\cdot\vec\tau_4=-3 .\label{31}
\end{eqnarray}
In this case, we obtain $E_{Excited}=23.658\,MeV$ by choosing the
value $M_{KK}=305 \,MeV$.

\subsection{Numerical Results}

In general, the considered potential in this model tends to zero at
$r\longrightarrow \infty$ and becomes infinity at small distances
which is well established for nuclear knowledge. The minimum of the
potential in the ground state is considered as the binding energy of
nucleus. Moreover, the difference between the minimum of the excited
state potential and the nucleus binding energy is considered as the
excited energy of the corresponding state. We apply our method for
the deuteron, $^{2}D$, Tritium, $^{3}T$ and two isotopes of Helium,
namely $^{3}He$ and $^{4}He$ nuclei.

To obtain the numerical results, $N_c=3$ have been chosen for the
realistic QCD. Also, we obtain the value of nucleon mass about
$m_N=0.92 \, GeV$ which is very close to the experimental nucleon
mass. In our Numerical calculations there is only one free parameter
$M_{KK}$.  The results of binding energy and excited energies are
compared with results of SS model and experiments in Tables VII and
VIII. As it is indicated from the tables, our results are in good
agreement with the experimental nuclear data. Moreover, our
potential has only one free parameters which allow us to fit our
results with the experimental data.

\begin{table}[htb]
\caption{\small The obtained binding energy of $^{2}D$, $^{3}T$,
$^{3}He$ and $^{4}He$ nuclei with $\Nc=3$ and $m_N=0.92 \, GeV$. The
results have a good consistency with the experimental nuclear data.
All energies are in $MeV$.} \center
\begin{tabular}{ccccc}
\hline \hline Nuclei   &  $ \quad M_{KK}\, $ & $\,\, E_B^{NC-H}
\,\,$& $\,\, E_B^{C-H}\,[51-52]\,$ & $\, E_{Exp}\,\, [89-92] \,$
\\ \hline
$^{2}D$  & 372  & 2.22    & 2.20   & 2.17 $\pm$0.0   \\
$^{3}T$  & 600  & 8.432   & 1.03   &  8.48        \\
$^{3}He$ & 372  & 7.8680  & 7.41   & 7.71         \\
$^{4}He$ & 533  & 28.3527 & 28.58  & 28.30        \\
\hline \hline
\end{tabular}
\end{table}

\begin{table}[htb]
\caption{\small The obtained excited energy of $^{3}He$ and $^{4}He$
nuclei with $N_c=3$ and $m_N=0.92 \, GeV$. The results have a good
agreement with the experimental nuclear data[89-90]. All energies
are in $MeV$.} \center
\begin{tabular}{cccccc}
\hline \hline $ Nuclei $  & $ J^P$  &  $M_{KK}$ & $ E_{Ex}^{NC-H} $&
$ E_{Ex}^{C-H}\,[18-19]\,$ & $\, E_{Ex}^{Exp}\,\, [89-90] \,$
\\ \hline
$^{3}He$ & $\frac{3}{2}^+$ &   -           & -         &    -       &  no state  \\
$^{4}He$ & $2^- , \, T=0$  &      395      & 21.8237   &   22.00    &  21.840    \\
$^{4}He$ & $2^- , \, T=1$  &      395      & 25.1001   &     -      &  23.330    \\
$^{4}He$ & $1^- , \, T=1$  &      305      & 23.658    &   23.17    &  23.640    \\
\hline \hline
\end{tabular}
\end{table}

In Table IX, we compare our numerical results for the light nuclei
binding energies with the predictions of the modern phenomenological
NN potential models [93]. It is obvious that our results obtained
using the non-critical holographic NN potential have a significant
agreement with the experimental data. It should be noted that we
calculated all of the parameters of noncritical holographic NN
potential [49] and also, our toy model for calculating the binding
energy have just one free parameter which is the mass scale of the
model, $M_{KK}$.

\begin{table}[htb]
\caption{\small 3N and 4N binding energies for various NN potentials
[93] compared with the our holographic model results and
experimental values. C-H and NC-H refer to the critical holographic
[39] and noncritical holographic potential [49] models,
respectively. All energies are in MeV.} \center
\begin{tabular}{cccc}
\hline
\hline  Potential  &  $ E_B(T)\, $   &$ E_B(^3He)\,$ &$ E_B(^4He)\, $  \\
\hline   CD Bonn   &   -8.012 &  -7.272  & -26.26  \\
AV18 &  -7.623  & -6.924 &   -24.28 \\
Nijm I &-7.736 & -7.085 & -24.98 \\
Nijm II & -7.654 & -7.012 & -24.56 \\
C-H  & -1.03 & -7.41 & -28.58  \\
NC-H & -8.4320 & -7.8680 & -28.3527  \\
\hline
Exp. & -8.48 & -7.72 &   -28.30\\
\hline \hline
\end{tabular}
\end{table}

Also, we compare our results for the $^{4}He$ binding energy with
the results obtained from other methods [94-95] such as
Faddeev-Yakubovsky (FY), Hyperspherical harmonics (HH), CCSD (CC
with singles and doubles), and  $\Lambda$-CCSD(T) (CC with triples
corrections ) in Table X. It is necessary to mention that our model
depends on just one parameter which is $M_{KK}$, whereas the other
theoretical models in nuclear literatures have more than one
parameters.

\begin{table}[htb]
\caption{\small Comparison of the $^4He$ binding energy obtained
from our model with the results of some other theoretical models
based on chiral low-momentum interactions [94-95].} \center
\begin{tabular}{cc}
\hline
\hline  Method   &  $ E_B(^4He)\, [MeV]$      \\
\hline
Faddeev-Yakubovsky (FY)                              &   - 28.65(5)  \\
 Hyperspherical harmonics (HH)                           &   - 28.65(2)  \\
 CCSD (CC with singles and doubles)                      &   - 28.44     \\
 $\Lambda$-CCSD(T) (CC with triples corrections ) &   - 28.63     \\
 Critical holography model (SS model)                    &   -28.58      \\
 Non-critical holography model($AdS_6$ model)            &   -28.3527    \\
\hline \hline
\end{tabular}
\end{table}


\section{Conclusion}

One of the applications of AdS/CFT correspondence is holography QCD
and introduced to solve the strong coupling QCD such the low-energy
dynamics of hadrons in particular baryons. A lot of holography
models are introduced to reproduce the QCD. Among them the SS model
is one of the most successful models due to its accurate results.
But, as we mentioned, the model encounters with some inconsistencies
in describing the baryons especially nucleons. For example, the mass
scale of the model to describe the nucleons are the half of the one
needs to describe the meson sector. Also, the size of baryon in the
large t'Hooft limit goes to zero. On the other hand, all holographic
QCD models based on the critical string theory suffer from the
unwanted KK modes.

In order to investigate these issues, we employ the noncritical
$AdS_6$ background and it's field theory dual. We study the mesons
and nucleons in this background and compute some of their features such
as the vector-meson spectrum, pion decay constant, baryon binding
energy, thermodynamic properties of baryonic matter, size of baryon,
nucleon-nucleon interaction and nucleon-meson coupling constants.

We review some obtained results in below which show that our results
not only are in a good agreement with the nuclear data, rather are
better than the SS model results.
\begin{enumerate}
    \item Just like the SS model, there exist some KK modes which come from the anti-periodic boundary conditions over the circle $S^1$. These
modes have the masses of the same order of magnitude as the lightest
glueballs of the four dimensional YM theory. Critical
holographic models such as the SS model, have some extra KK modes
too which do not belong to the spectrum of pure YM theory. These
undesired KK modes come from the extra internal space over which ten
dimensional string theory is compactified, for example, the $S^4$
sphere in the SS model. In the non-critical holographic model, which
we used here, there is no additional compactified sphere, so there
are no such extra KK modes and the QCD spectrum is clear from them.
Thus it seems that our model based on the non-critical holography is
much more reliable.

    \item We studied the dynamics of gauge field living on the
flavor probe brane and obtained the spectrum of vector mesons. Our
results were compared with the result of other holographic models and
the experimental data. Also, we calculated the pion decay constant
in terms of model parameter. We found the values of mass scale
$\Mkk=395 \,MeV$ to have pion decay constant $f_{\pi}=92\,MeV$.

    \item In order to study the nuclear physics in the holography
    frame, we investigated baryons which are defined by a vertex
    with $\Nc$ fundamental strings attached to the flavor brane. We
    obtained the binding energy of baryon in the noncritical $AdS_6$
    model [31]. Baryon in holography is replaced by a solitonic
    instanton such that the instantonic number shows the baryon
    number. We used this definition of baryon in the $AdS_6$ model
    and calculated it's size. We demonstrated that the size of
    baryon is of order one, therefor the zero size of baryon in the
    holography SS model was solved here [49].

    \item Holographic models have a mass scale which is the
    low-energy scale of the model. In the SS model, the value of
    $\Mkk$ to describe the baryon should be half of one to
    describe the mesons. The nucleon mass was obtained roughly $920\, MeV$ using
    $\Mkk=395\,MeV$. So, our model could describe the mesons and
    nucleons with the same mass scale well.

    \item We employed the noncritical $AdS_6$ model to study the
    NN potential and nucleon-meson coupling constants. We derived
    the Yukawa coupling constants by exploring the dynamics of
    nucleon in the holography frame. We compared our results with
    the predictions of four modern phenomenological NN potential
    models. The remarkable point is that all nucleon-meson coupling
    constants have been calculated in the holography model, whereas
    these parameters were obtained by fit to the empirical NN
    scattering data in the phenomenological potentials. Our
    holography NN potential can be more accurate by considering the
    derivative couplings in the magnetic channels. In addition, the
    holography NN potential obtained using the $AdS_6$ model, can be
    used widely in describing the nuclear structure and
    multi-nucleon systems such as the nuclear binding energy and NN
    scattering.

    \item The small value of nuclear binding energy is one of the
    interesting issues in nuclear physics. Despite of the power of strong interaction, the NN force is
    small: binding energy is only a few percent of the mass of the
    nucleons.
    In the holographic models, the exchange of heavy mesons are
    suppressed in the large $\Nc$ limit. As a result, the
    interaction of two nucleons is explained via the exchange of
    light mesons such as pion and $\omega$-meson. The exchange of
    pion is responsible to the attractive long-range nuclear force.
    Whereas, the exchange of $\omega$-meson produce mainly
    medium-range repulsive force. If we suppose the repulsion starts
    at distance $|x|\sim m_{\omega}^{-1}$, then the nuclear binding
    energy is of order $E_{binding} \sim \frac{1}{g_s} m_{\omega}$  which
    is much smaller than the nucleon mass. Above analysis motivated
    us to introduce a simple toy model to estimate the binding
    energy of multi-nucleons systems. We explained the model in
    the previous section in details. In general, the obtained nuclear potential
    have the familiar behavior in nuclear physics. In addition,
    despite of the small number of free parameters in our holography
    model, the obtained results have significant agreement with the
    experimental data.

    \item In our holography model for the light nuclei, we assumed
    that the setting of a small number of instantonic D-brane on the
    background does not change the background. In fact, we ignored
    the backreaction of baryon vertices and background geometry. It
    is clear that this assumption is correct just for the light
    nuclei. In fact, one can find a gravity dual for heavy nuclei by
    implying the AdS/CFT correspondence again. In this holographic
    description, the gauge theory on the nuclei with mass number A,
    is U(A). Study of the U(A) gauge theory is hard, but the theory
    becomes more simple by taking the large A limit. In this limit,
    one can find the near horizon geometry dual to the gauge theory.
    The supergravity solution has a discrete spectrum which is the
    excited spectrum of heavy nuclei with mass A [17]. The result is in
    agreement with nuclear data manifestly.
    As we know from the nuclear experiments, the nucleons of a heavy
    nuclei have coherent excitations which are called Giant
    resonances. These resonances exhibit harmonic behavior $E_n=n\,
    w(A)$ which is explained with phenomenological models such
    as the liquid drop model. The gauge-gravity duality can
    reproduce this behavior. Also, dependence to the mass number A
    is obtained by using the duality [17].

    In this regard, several issues can be studied. For example, if
    we put a probe brane as an external nucleon near the
    near-horizon geometry of A D-brane and consider the probe
    dynamics, the shell model potential of nuclear physics may
    obtained.

    On the other hand, since blackholes are described by fluid
    dynamics holographically, one can speculate that the liquid drop
    model of heavy nuclei may be related to dual geometries through
    the holographic hydrodynamics. In fact, dissipation of
    excitations on a nucleus is a target of research for many
    decades.

    \item The repulsive core potential is one of the critical issues
    of nuclear physics that its origin is still not well
    understand. Nuclear force has been studied using the AdS/CFT correspondence [13-16]
    and an explicit expression has been obtained for the nuclear
    force which contains the repulsive core too. This potential
    behaves as $r^{-2}$ in small distances. However, there are a lot of
    unanswered questions about the nuclear repulsive and attractive force yet.

\end{enumerate}

Finally, it seems that the AdS/CFT correspondence is a new tool to
solve the unanswered questions in nuclear physics.


\end{document}